\documentclass[11pt]{article}
\usepackage{graphicx,color}  
\usepackage{dcolumn}   
\usepackage{bm}        
\usepackage{amssymb} 
\usepackage{amsmath}
\usepackage{braket}
\usepackage{cancel}
\usepackage{subfigure}
\usepackage{fullpage}
\newcommand{\ka}{\kappa}
\newcommand{\del}{\delta}
\newcommand{\lag}{\langle}
\newcommand{\rag}{\rangle}
\newcommand{\pa}{\partial}
\newcommand{\cA}{\mathcal{A}}

\def\a{\alpha}
\def\b{\beta}

\def\m{\mu}
\def\n{\nu}
 \let\h=\eta

\newcommand{\cS}{\mathcal{S}}
 \newcommand{\beq}{\begin{equation}}
\newcommand{\eeq}{\end{equation}}
\newcommand{\beqa}{\begin{eqnarray}}
\newcommand{\eeqa}{\end{eqnarray}}
\newcommand{\be}{\begin{equation}}
\newcommand{\ee}{\end{equation}}
\newcommand{\bea}{\begin{eqnarray}}
\newcommand{\eea}{\end{eqnarray}}

\newcommand{\sdfrac}[2]{\mbox{\small$\displaystyle\frac{#1}{#2}$}}

\newcommand{\nc}{\newcommand}
\nc{\rnc}{\renewcommand}
\nc{\D}{\partial}
\nc{\K}{\kappa}
\nc{\bK}{\bar{\K}}
\nc{\bN}{\bar{N}}
\nc{\bq}{\bar{q}}
\nc{\vbq}{\vec{\bar{q}}}
\nc{\g}{\gamma}
\nc{\lrarrow}{\leftrightarrow}
\nc{\rg}{\sqrt{g}}
\nc{\de}{\delta}
\let\d=\delta
\nc{\nn}{\nonumber}
\rnc{\(}{\left(}
\rnc{\)}{\right)}
\nc{\q}{\vec{q}}
\nc{\x}{\vec{x}}
\nc{\ep}{\epsilon}
\nc{\tto}{\rightarrow}
\rnc{\inf}{\infty}
\rnc{\Re}{\mathrm{Re}}
\rnc{\Im}{\mathrm{Im}}
\nc{\z}{\zeta}
\nc{\mA}{\mathcal{A}}
\nc{\mB}{\mathcal{B}}
\nc{\mC}{\mathcal{C}}
\nc{\mD}{\mathcal{D}}
\nc{\mN}{\mathcal{N}}
\rnc{\H}{\mathcal{H}}
\rnc{\L}{\mathcal{L}}
\nc{\<}{\langle}
\rnc{\>}{\rangle}
\nc{\fnl}{f_{NL}}
\nc{\gnl}{g_{NL}}
\nc{\fnleq}{f_{NL}^{equil.}}
\nc{\fnlloc}{f_{NL}^{local}}
\nc{\vphi}{\varphi}
\nc{\Lie}{\pounds}
\nc{\half}{\frac{1}{2}}
\nc{\bOmega}{\bar{\Omega}}
\nc{\bLambda}{\bar{\Lambda}}
\nc{\dN}{\delta N}
\nc{\gYM}{g_{\mathrm{YM}}}
\nc{\geff}{g_{\mathrm{eff}}}
\nc{\tr}{\mathrm{tr}}
\nc{\oa}{\stackrel{\leftrightarrow}}
\nc{\IR}{{\rm IR}}
\nc{\UV}{{\rm UV}}
\nc{\la}{\lambda}
\nc{\veps}{\varepsilon}

\allowdisplaybreaks

\begin{document}
	\begin{center}
		\vspace{1.5cm}
		{\Large \bf CFT Correlators and CP-Violating Trace Anomalies \\}
		\vspace{0.3cm}
		
		\vspace{0.3cm}
		
		\vspace{1cm}
		{\bf Claudio Corian\`o$^{(1)}$, Stefano Lionetti\bf $^{(1)}$ and Matteo Maria Maglio$^{(2)}$ \\}
		\vspace{1cm}
		{\it  $^{(1)}$Dipartimento di Matematica e Fisica, Universit\`{a} del Salento \\
			and INFN Sezione di Lecce, Via Arnesano 73100 Lecce, Italy\\
			National Center for HPC, Big Data and Quantum Computing\\}
		\vspace{0.5cm}
		{\it  $^{(2)}$ Institute for Theoretical Physics (ITP), University of Heidelberg\\
			Philosophenweg 16, 69120 Heidelberg, Germany}

		\begin{abstract}
		We analyze the parity-odd correlators $\langle JJO\rangle_{odd}$, $\langle JJT\rangle_{odd}$, $\langle TTO\rangle_{odd}$ and $\langle TTT\rangle_{odd}$ in momentum space, constrained by conformal Ward identities, extending our former investigation of the parity-odd chiral anomaly vertex. We investigate how the presence of parity-odd trace anomalies affect such correlators. Motivations for this study come from holography, early universe cosmology and from a recent debate on the chiral trace anomaly of a Weyl fermion. In the current CFT analysis, $O$ can be either a scalar or a pseudoscalar operator and it can be identified with the trace of the stress-energy tensor. We find that the $\langle JJO\rangle_{odd}$ and $\langle TTO\rangle_{odd}$ can be different from zero in a CFT. This can occur, for example, when the conformal dimension of the scalar operator is $\Delta_3=4$, as in the case of $O=T^\mu_\mu$. Moreover, if we assume the existence of parity-odd trace anomalies,
		 the conformal $\langle JJT\rangle_{odd}$ and $\langle TTT\rangle_{odd}$ are nonzero. In particular, in the case of $\langle JJT\rangle_{odd}$ the transverse-traceless component is constrained to vanish, and the correlator is determined only by the trace part with the anomaly pole. 
 		\end{abstract}
	\end{center}
	\titlepage

\section{Introduction}

Among the quantum field theory anomalies, the chiral anomaly is one of the most discussed, both in its global and gauged versions. It appears in parity-odd correlators with multiple vectors $(J)$ and axial vector $(J_5)$ currents, usually denoted as the $\langle AVV\rangle$ and $\langle AAA\rangle$, or in other parity-odd correlators such as the $\langle J_5 TT\rangle$, where an axial vector current  is contracted with two stress-energy tensors $(T)$. The anomalous conservation identity is
\begin{equation} \label{eq:anomaliachirale}
	\nabla_\mu\langle J_5^\mu\rangle =
	a_1\, \varepsilon^{\mu \nu \rho \sigma}F_{\mu\nu}F_{\rho\sigma}+ a_2 \, \varepsilon^{\mu \nu \rho \sigma}R^{\alpha\beta}_{\hspace{0.3cm} \mu \nu} R_{\alpha\beta \rho \sigma}.
\end{equation}
The anomalous terms on the right-hand side are called Pontryagin densities.
The first term ($F\tilde F$) appears in the analysis of the $\langle AVV\rangle$ and $\langle AAA \rangle$ correlators while the second one ($R\tilde{R}$) is encountered in the analysis of the $\langle J_5TT\rangle $ correlator. \\
Recently, we have investigated the $\langle AVV\rangle$ and $\langle AAA \rangle$ correlators in the context of CFT in momentum space \cite{Coriano:2023hts}. We have shown how the expression of such correlators can be reconstructed by requiring that they satisfy conformal Ward identities (CWIs). The approach avoids any reference to a perturbative realization and is, in this respect, nonperturbative. We only require, as boundary condition on the solution of the differential equations, that the divergences of the 
$J_5$ and $J$ currents are either anomalous or conserved. In other words, CFT does not tell us whether a certain correlator is anomalous or not, but once we set the anomaly boundary conditions on their operators, motivated by physical considerations or by previous perturbative analysis, the CWIs - anomalous or ordinary - constrain significantly its expression.   \\
For example, if we separate the $\langle AVV\rangle $ into a longitudinal and a transverse part, the CWIs are then sufficient to determine its transverse part explicitly, once its longitudinal part is constrained by the anomaly. The boundary condition on the divergence of the current is solved by introducing a minimal, specific,  tensor structure, containing an anomaly pole. In this paper, we review very briefly the procedure followed in the case of the anomalous $\langle AVV \rangle $ correlator in appendix \ref{appendix:AVV}, that may help to clarify the approach. The final expression of the correlator that we obtain depend on the parameter that constrains its longitudinal structure. \\
For the parity-even case, one can recover, for instance, Furry's theorem on the vanishing of the $\langle VVV\rangle $ vertex, without directly invoking charge conjugation (C) invariance, as in the perturbative analysis. The only conditions that one has to impose on the correlator are just the conservation of the three vector currents and the conformal Ward identities. \\
 \subsection{The inclusion of the anomaly}
 For the moment we just recall that perturbative analysis of parity-odd correlators reveals that an anomaly, either chiral or conformal, as just mentioned, is always associated with the presence either of a longitudinal or of a trace structure, in both cases  characterised by the inclusion of an anomaly pole 
\cite{2009PhLB..682..322A, Giannotti:2008cv}. 
The pole term, as identified by the longitudinal/transverse decomposition \cite{Knecht:2003xy} in the $\langle AVV\rangle$, takes the role of a pivot for the determination also of all the form factors of this diagram by imposing the CWIs, i.e. the remaining transverse ones.  
In this way, a single longitudinal tensor structure and form factor accounts for the anomaly, enforced by external anomalous (for chiral currents) and ordinary (for the vector currents) conservation Ward identities, while the transverse part is obtained by solving the conformal constraints. For the $\langle AVV\rangle$ correlator, the two parts of the decomposition are coupled together by the conformal equations. \\

\subsection{Parity-odd trace anomalies and regularizations}
Conformal anomalies, on the other end, have some distinctive features that make them more complex compared to the chiral ones, due to the presence of both topological and of non topological terms in the anomaly functional. We recall that trace anomalies are associated with the generation of a parity-even anomaly functional, given by 
\beq
\label{trace1}
g_{\mu\nu}\langle T^{\mu\nu}\rangle =  b_1\, {E}_4+b_2 \, C^{\mu \nu \rho \sigma} C_{\mu \nu \rho \sigma}+b_3\, \nabla^2 R+b_4\, F^{\mu\nu}F_{\mu\nu},
\eeq
where $C_{\mu \nu \rho \sigma}$ is the Weyl tensor and ${E}_4$ is the Gauss-Bonnet term
\begin{equation}
	\begin{aligned}
		C^{\mu \nu \rho \sigma} C_{\mu \nu \rho \sigma} & =R^{\mu \nu \rho \sigma} R_{\mu \nu \rho \sigma}-2 R^{\mu \nu} R_{\mu \nu}+\frac{1}{3} R^2, \\
		{E}_4 & 
		\equiv E=R^{\mu \nu \rho \sigma} R_{\mu \nu \rho \sigma}-4 R^{\mu \nu} R_{\mu \nu}+R^2.
	\end{aligned}
\end{equation}
However, as was
first found by Capper and Duff \cite{Capper:1974ic,Duff:2020dqb} on dimensional grounds and by requiring covariance, the structure of the trace anomaly in four dimensions can be more general than \eqref{trace1} and constrained to be of the form
\begin{equation}
\label{ann}
	\mathcal{A}= b_1\, {E}_4+b_2 \, C^{\mu \nu \rho \sigma} C_{\mu \nu \rho \sigma}+b_3\, \nabla^2 R+b_4\, F^{\mu\nu}F_{\mu\nu}+f_1\, \varepsilon^{\mu \nu \rho \sigma} R_{\alpha \beta \mu \nu} R_{\,\,\,\,\, \rho \sigma}^{\alpha \beta}+f_2\,\varepsilon^{\mu\nu\rho\sigma}F_{\mu\nu}F_{\rho\sigma},
\end{equation}
which includes both parity-even and parity-odd terms. \\
The parity-odd ones were investigated in the action in \cite{Deser:1980kc} as possible sources of CP violation induced by gravity, providing a possible solution to a long standing problem.  
Also, their connection with the anomalies was central in the investigation of the quantum inequivalence of different representations of antisymmetric tensor fields  coupled to gravity \cite{Duff:1980qv}, \cite{Sezgin:1980tp}. Therefore, there are very significant reasons to 
investigate their role in a clear physical context, especially in a phase of the early universe where conformal symmetry is expected to play a fundamental role.\\
Under parity inversion, all the terms in the trace anomaly are invariant except for the last two of them. The parity of the $F\tilde F$ term is odd, while for the $R\tilde{R}$ it is not naturally defined in a curved background. For both terms, though, we expect the 
coefficients to be real in order to comply with unitarity. We recall that 
$F\tilde{F}\sim E\cdot B$ is CP-odd as well as time reversal odd. Indeed, if we consider that the stress-energy tensor is a fundamental composite operator of the Standard Model (SM), the presence of imaginary coefficients would endanger the consistency of the theory.
All the coefficients  in eq. \eqref{ann} have been computed in the parity-even case, and their values strictly depend on the number and type of massless fields entering in the perturbative quantum corrections, but they are real. 
The computation of the parity-odd ones has also been performed in free field theory realizations in different regularization schemes, from dimensional reduction (DRED) \cite{Armillis:2010pa} to Pauli Villars (PV)  \cite{Bastianelli:2016nuf,Bastianelli:2018osv} to the 't Hooft-Veltman Breitenlohner-Mason (HVBM) scheme \cite{Abdallah:2021eii}, showing that they are zero. \\
Without entering into the debate whether free field theory, and the SM in particular, is affected by such parity-odd anomaly terms, one 
may ask, on general grounds, whether CFTs allow such parity-odd terms in the case in which the $f_i$'s are generic, and the possible constraints on the structure of the correlators. Their real or imaginary character is not relevant for our purposes and is independent  of the fact that a free field theory realization is possible. We are going to find significant constraints on their structure from the solution of their CWIs.

\subsection{Motivations and methodology}
As mentioned, several computations of these correlators with Weyl fermions in different regularization schemes indicate that the $f_i$ vanish.
An older previous analysis in free field theory of the $\langle TJJ\rangle_{odd}$, 
reached this conclusion \cite{Armillis:2010pa}. Several further studies have confirmed this result \cite{Bastianelli:2016nuf,Bastianelli:2018osv,Bastianelli:2019fot,Bastianelli:2019zrq,Frob:2019dgf,Abdallah:2021eii,Bastianelli:2022hmu,Abdallah:2023cdw}.
In the last few years, however, following the work of \cite{Nakayama:2012gu}, other groups have claimed the existence of non-vanishing parity-odd terms in the trace anomaly of Weyl fermions, employing either a diagrammatic approach or the heat-kernel method \cite{Bonora:2014qla,Bonora:2015nqa,Bonora:2017gzz,Bonora:2020upc,Bonora:2021mir,Bonora:2022izj}. Moreover, new independent results have appeared recently, in support of these alternative conclusions \cite{Liu:2022jxz,Liu:2023yhh} (see \cite{Nakagawa:2020gqc,Nakagawa:2021wqh} for the supersymmetric version). Notice that in these computations the coefficients $f_i$ are found to be nonzero and imaginary, raising a crucial issue with unitarity. 
These results, if confirmed, would be of remarkable phenomenological relevance, since such anomalies would prove that the chiral spectrum of the Standard Model, for its consistent coupling to gravity, needs to be modified with the addition, for example, of extra chiral fermions. For example, if we accept these results, we can prove quite immediately that the $\langle T J_Y J_Y \rangle $, where $J_Y$ is the hypercharge current of the Standard Model (SM), does not vanish and extra chiral matter is needed.\\
The study of parity odd trace anomalies is related also to the holographic contest.
Indeed, in \cite{Nakayama:2012gu,Nakayama:2013oqa} the author presented a holographic model which yields  a Pontryagin density in the trace of the e. m. tensor, but again with unitarity problems.
\\
Our analysis, here, will be limited to abelian gauge currents and parity-odd anomalies of 3-graviton vertices 
$(TTT)$, leaving further, more general studies of other correlators to future work. \\
The use of the CWIs to determine parity-odd anomaly correlators - independently of any free field theory realization - has been first discussed in \cite{Coriano:2023hts} in the chiral anomaly case. Here we are going to extend that analysis, finding very strong constraints on the correlators that we investigate. \\
Concerning the methodology that we use, we recall that the general solution of the CWIs have been successfully analyzed in momentum space \cite{Coriano:2013jba}  \cite{Bzowski:2013sza}  in the scalar and tensor cases respectively, for parity-even correlators, in the presence of ordinary (parity-even) trace anomalies, corresponding the the Euler-Poincar\`e density $E$ and the square of the Weyl tensor $C$.\\
 In this paper, we follow closely the procedure adopted in \cite{Bzowski:2013sza}, and investigate the CWIs using a similar decomposition of the correlators into transverse traceless, trace and longitudinal sectors. 
 We will investigate the $\langle JJO\rangle $, $\langle JJT\rangle $ and $\langle TTO\rangle $ correlators in their CP-odd sectors, 
 while in the case of the $\langle TTT \rangle $ our analysis will be limited to the longitudinal and trace sectors, leaving the transverse-traceless sector to incoming work. This sector of the $\langle TTT \rangle $ is less relevant for the purpose of a comparison with the perturbative prediction and the ongoing debate over the vanishing or non vanishing of the parity-odd parts of such correlator, since 
no parity-odd anomaly has been predicted for the $\langle TTT\rangle $ in all the computations. 
This is the case if the trace anomaly is defined in the usual sense, as a trace operation performed {\em after} the computation of the quantum corrections as
\beq  
\label{def}
\mathcal{A}_{odd}\equiv g_{\mu\nu} \langle T^{\mu\nu}\rangle. 
\eeq
In this case, then the $\langle TTT\rangle_{odd}$ can be nonzero only if we allow a nonzero trace anomaly in the correlators obtained by the functional differentiation of \eqref{def}, otherwise it is identically zero. 
In the perturbative analysis of  \cite{Bonora:2014qla,Bonora:2015nqa,Bonora:2017gzz,Bonora:2020upc,Bonora:2021mir,Bonora:2022izj}, the anomaly - in the case of the $\langle TTT\rangle $ -  comes instead from the second (subtraction) term in the modified definition of the functional $\mathcal{A}$ according to \eqref{eq:defanomduff}. This is a subtraction term in which the trace is performed before the quantum average.  In our CFT analysis, as we are going to explain, the subtraction term corresponds to the computation of  the anomaly of the $\langle TTO\rangle$ and $\langle JJO\rangle$, with the scalar or pseudoscalar operator $O$ identified with the trace of $T$.\\
Our results show some interesting features of the correlators, quite different in the $\langle JJT\rangle $ and $\langle TTT\rangle $ cases. For example, we show that parity-odd correlators such as the $\langle JJT\rangle$ or the $\langle JJO\rangle$ and  $\langle TTO\rangle$ all satisfy non anomalous CWIs. The only correlator that satisfies anomalous CWIs is the $\langle TTT \rangle$ limitedly to the special CWIs.\\
From this perspective, the result is quite similar to the case of the $\langle AVV\rangle $ chiral anomaly correlator, that also satisfies ordinary special and dilatation CWIs. Indeed, in both cases, in the perturbative relizations of such correlator, there is no counterterm available, since the $F\tilde F$ and $R\tilde R$ terms are both topological and all the form factors become finite, just by enforcing the conservation WIs on the vector currents. \\
In the case of the $\langle JJT\rangle$, the correlator can be nonvanishing if we allow an anomalous trace as in \eqref{def}. The interesting fact is that the special and dilatation CWIs are non anomalous and the solution satisfies the ordinary conformal constraints even in the presence of a nonzero trace. This solution is anyhow rather peculiar, since all the other sectors of this correlator (transverse-traceless and longitudinal) are constrained by the same equations to vanish. \\
We recall that, recently, novel approaches have been adopted for the construction of 
parity-odd correlation functions. In particular in \cite{Jain:2021gwa} it has been shown that 
parity-odd CFT 3-point functions can be obtained by acting in a specific way on the 
parity-even sector of some correlation functions. Concerning the requirement of dealing with non-conserved currents, these have been investigated, for instance, in  \cite{Marotta:2022jrp} even for higher spin operators, but only in the parity-even case. 
Moreover, in \cite{Jain:2021wyn} the authors use both the momentum space CWIs as well as spin-raising and weight-shifting operators to fix the form of parity-odd correlators. However, both \cite{Jain:2021gwa} and \cite{Jain:2021wyn} investigate anomaly-free cases.

\section{Definition of the anomaly and scheme dependence}

If an anomaly is interpreted as the failure of the trace operation to commute with a quantum average, then the trace anomaly may also be defined as the difference of two trace operations on the stress-energy tensor, one performed before the quantum average and one after, as proposed by Duff  \cite{Duff:1993wm,Duff:2020dqb}
\begin{equation} \label{eq:defanomduff}
	\mathcal{A}=g^{\mu \nu}(x)\left\langle T_{\mu \nu}(x)\right\rangle-\left\langle T_\mu^\mu(x)\right\rangle,
\end{equation}
which is the definition considered in \cite{Bonora:2022izj}. \\
We pause to define our notations.
In the following we will denote with $O\equiv T^\mu_\mu$ the trace of the stress-energy tensor, while $T$ will be denoting the same operator but with free indices. In the case in which the stress-energy tensor is parity-odd, its trace 
is a pseudoscalar operator, at times indicated as $O_5\equiv {T_5}^{\mu}_{ \mu}$, whenever necessary. The notation is reminiscent of the ways in which the $\gamma_5$ chiral projector is positioned in the loop of the Feynman expansion, either at the $T$ vertex or at the $J$ vertex, turning a $T$ into a $T_5$ or a $J$ into a $J_5$. 
In the context of correlators containing CP-odd operators of a CFT, is may be useful to characterize which operators are CP-even and which are CP-odd, since this information is crucial for the consistency of the WIs. For instance, parity-odd 2-point functions $\langle JJ \rangle_{odd}$ vanish, since one of the two $J$'s is actually a $J_5$. \\
Therefore, in the parity even case \eqref{eq:defanomduff} can then be written in the form 
\begin{equation} \label{eq:defanomduff1}
	\mathcal{A}=g^{\mu \nu}(x)\left\langle T_{ \mu \nu}(x)\right\rangle-\left\langle O(x)\right\rangle,
\end{equation}
and in the parity-odd case as 

\begin{equation} \label{eq:defanomduff2}
	\mathcal{A}=g^{\mu \nu}(x)\left\langle T_{5 \mu \nu}(x)\right\rangle-\left\langle O_5(x)\right\rangle.
\end{equation}
However, we will try to avoid any cumbersone notation unless it is strictly necessary. 
The importance  of \eqref{eq:defanomduff2} is that, for a non conformal theory, the contributions to the trace which are non invariant under the symmetry - for instance mass dependent terms - are removed by the subtraction. 
However, according to Bonora \cite{Bonora:2022izj}, the last term in eq. \eqref{eq:defanomduff} gives also nonvanishing contributions to the trace anomaly in the massless limit.\\
A peculiar character of the regularization discussed in \cite{Bonora:2022izj} is that a term which was initially introduced to cancel classical contributions in non-conformal theories - the second term of \eqref{eq:defanomduff2} - can be responsible for generating the entire parity-odd trace anomaly $R \tilde{R}$.\\
Moreover, the results seem to depend on how the scheme and regularization prescriptions are applied. 
However, as suggested in \cite{Bonora:2022izj},
one may get different perturbative results because one stops at the lowest order. If one was able to work out the next perturbative orders, while still preserving diffeomorphism covariance, one would find the same results (0 or not 0), independently of the regularization scheme.
\\
In schemes based on dimensional reduction DRED, where the traces are performed in four dimensions, the perturbative computations are quite simple. In this regularization scheme, in 3-point functions containing a parity-odd sector, the first term can be defined in multiple but equivalent ways - i.e. for the case of gauge contributions $F\tilde{F}$ by tracing either $\langle JJT_5 \rangle$ or $ \langle J_5 J T\rangle$ or $\langle  J_5 J_5 T_5\rangle$ - all of them characterised, in the perturbative analysis, by an odd number of $\gamma_5$. We classify these cases, from the CFT viewpoint, simply as $\langle JJT\rangle_{odd}$.\\
Similarly, the second term can be equivalently defined as  $\langle JJO_5\rangle$ or $\langle  J_5 JO\rangle $ or $\langle  J_5 J_5O_5\rangle$, since the position of the projector is irrelevant in the identification of the transverse traceless sector of this correlator. We classify these cases from the CFT viewpoint simply as $\langle JJO\rangle_{odd}$. As already mentioned, in the hierarchical WIs, some care is however necessary. \\
We will come back to address the connection between our results and perturbation 
theory in a separate work. Here we will simply focus on the solution of the CFT constraints on such correlators coming from the CWIs in the presence of parity-odd traces, with an anomaly defined as in \eqref{eq:defanomduff}. 
  
\subsection{Anomalous CWIs and the absence of a $0/0$ limit }
\label{anom1}
The correlation functions are fixed by anomalous CWIs that can be written down directly at $d=4$. In general, 
the method consists in deriving the ordinary (i.e. non anomalous) CWIs satisfied by the theory in general $d$ dimensions, and introducing a renormalization procedure that allows to remove the divergences generated by the solution in the $d\to 4$ limit. This approach has been developed, independently of any free field theory realization, in \cite{Bzowski:2013sza,Bzowski:2017poo, Bzowski:2018fql} by introducing suitable counterterms for each type of correlator. 
In the case of parity even trace anomalies, correlators such as the $\langle TJJ\rangle_{even}$ satisfy ordinary CWIs in $d$ dimensions and are renormalized by the inclusion of a ${1}/({d-4})\,F^2$ counterterm. Obviously, the types of counterterms needed in the renormalization of a generic CFT are those that appear in the expression of the trace anomaly in the parity even case. \\
For correlators such as the 
$\langle TTT\rangle_{even}$ two counterterms are needed, $ V_E/\epsilon$ and $ V_{C^2}/\epsilon$, with  
\begin{align}
\label{ffr}
V_{E}(g,d)\equiv &\mu^{\epsilon} \int\,d^dx\,\sqrt{-g}\,E, \qquad \qquad \epsilon=d-4\notag \\ 
V_{C^2}(g, d )\equiv & \mu^{\epsilon}\int\,d^dx\,\sqrt{-g}\, C^2. 
\end{align}
$E$ is evanescent, for being topological at $d=4$, but it is necessary in order to satisfy in the anomaly effective action, the Wess-Zumino consistency condition. The generation of the anomaly, in DR, is 
associated with the non invariance of the two counterterms under Weyl variations 
\begin{equation} \label{ren}
	\begin{aligned}
		&2 g_{\mu\nu}\frac{\delta}{\delta g_{\mu\nu}}V_{E}=\frac{\delta}{\delta \phi}V_{E}=\epsilon \sqrt{g} {E},\\&
		2 g_{\mu\nu}\frac{\delta}{\delta g_{\mu\nu}}V_{C^2}=\frac{\delta}{\delta \phi}V_{C^2}=\epsilon \sqrt{g}
		\bigg[ C^2+\frac{2}{3} \square R \bigg]
	\end{aligned}
\end{equation}
where we have introduced the conformal decomposition of the metric  
\beq
g_{\mu\nu}=\bar{g}_{\mu\nu}e^{2 \phi}
\eeq
in terms of a conformal factor $(\phi)$ and a fiducial metric $\bar{g}$. Notice that in the case of a 
topological contribution $(E)$ the extension of the $\varepsilon^{\mu\nu\rho\sigma}$  tensor is a matter 
of prescription. In the parity-odd case, terms such as $R\tilde{R}$ and $F\tilde{F}$ cannot really play the role of counterterms in the derivation of a trace anomaly - in this case inducing a finite renormalization of the quantum corrections -  since they are not linear in $\epsilon$ under Weyl variations and we cannot perform the $\epsilon\to 0$ limit. We recall that in DR, the evanescence is related to the $0/0$ behaviour of a certain counterterm, which is clearly ensured by the relations \eqref{ren} but not in the parity-odd case.
 Therefore it is certainly true, from these types of considerations, that the singularities of the corresponding diagrams should disappear once we impose some external WIs on the specific correlator. 
However, topological anomalies can be generated also without resorting to any regularization procedure, such as in the $\langle AVV\rangle$ chiral anomaly diagram. \\

\section{Anomalous CWIs and their parity-odd extension} 
In this section we examine the anomalous CWIs relying on a derivation that is strictly enforced 
at $d=4$, using a geometric approach. This allows us to overlook for a moment at the nature of the $R\tilde R$ and $F\tilde{F}$ counterterms, as we have just mentioned above, by assuming that both terms are present in the trace anomaly. \\
It is convenient - and also more direct - to follow the derivation of these equations starting from the functional integral defining the partition function via the anomaly action $\cS$
\be
\exp \big\{i\, \cS[g]\big\} \equiv \int [d \Phi] \exp\big\{i\, {\rm S_{cl}}[\Phi, g]\big\}
\label{Sexact}
\ee
over all the matter/radiation fields (at all scales), here denoted as $\Phi$, since the equations that we obtain can be naturally extended to the correlators of a non-Lagrangian CFT.  
By expanding this 1PI quantum action functional in a Taylor series
\bea
&&\cS[g+h] = \cS[g] + \\
&&\hspace{-5mm}\sum_{n=1}^{\infty} \frac{1}{2^n\,n!} \int d^4x_1\dots d^4x_n \, \sqrt{-g(x_1)} \dots \sqrt{-g(x_n)}\, \cS^{\m_1\n_1\dots \m_n\n_n}_n(x_1, \dots,x_n;g)\,
h_{\m_1\n_1}(x_1)\dots h_{\m_n\n_n}(x_n),   \nonumber
\label{Tayexp}
\eea
we define the $n-$point coefficients of the expansions as
\be
\cS_n^{\m_1\n_1\dots\m_n\n_n}(x_1,\ldots,x_n;g) \equiv \frac{2^n}{\sqrt{-g(x_1)}\dots\sqrt{-g(x_n)}}\ 
\frac{\del^{n}\cS[g]}{\del g_{\m_1\n_1}(x_1)\dots \del g_{\m_n\n_n}(x_n)}
\label{Tn}
\ee
that equal the correlation functions of stress-energy tensors for non-coincident spacetime points 
\beq
\langle T^{\mu_1\nu_1}(x_1)\ldots T^{\mu_n\nu_n}(x_n)\rangle\equiv \cS_n^{\mu_1\nu_1\ldots \mu_n\nu_n}(x_1\ldots x_n)
\eeq
in a metric background $g$. 
For $n=1$ the one-point function
\be
\cS^{\m\n}_1(x;g) \equiv \big\lag T^{\m\n}(x)\big\rag_g=\frac{2}{\sqrt{-g(x)}}\ \frac{\del \cS [g]}{\del g_{\m\n}(x)} 
\label{T1def}
\ee
equals the renormalized expectation value of $T^{\m\n}(x)$ and $\cS [g]$ is the finite renormalized 1PI effective action. \\
In the case of a pseudoscalar operator $O$, the parameterization of the correlation functions derived from its quantum average will be performed as usually done in the longitudinal/transverse decomposition, in full generality. The contribution of such correlation functions to the CWIs, will be derived by observing that the subtraction term in \eqref{eq:defanomduff2} will be defined using the operation 
\beq
\langle O(x_1)\rangle =\langle 2 g_{\mu\nu}(x_1)\frac{\delta }{\delta g_{\mu\nu}(x_1)} {\rm S_{cl}} \rangle,
\label{onec}
\eeq
performed before any further differentiation with repect to the background metric

\beq
\langle T T O\rangle =\frac{2}{\sqrt{{g}(x_1)}}\frac{2}{\sqrt{{g}(x_2)}}\frac{\delta}{\delta g_{\mu_1\nu_1}(x_1)}\frac{\delta}{\delta g_{\mu_2\nu_2}(x_2)} \langle O(x_1)\rangle .
\label{twoc}
\eeq
Let us now introduce the following quantity which we are going to use in the next sections
\begin{equation}\label{eq:atildeaa}
	\mathcal{\tilde{A}}\equiv2 g_{\mu\nu}\frac{\delta \mathcal{S}[g]}{\delta g_{\mu\nu}}=\sqrt{g}\, g_{\mu\nu}T^{\mu\nu}= \, f_1 \sqrt{g}\, \varepsilon^{\mu\nu\rho\sigma} R_{\mu\nu}^{\,\,\,\,\,\,\alpha\beta}R_{\rho\sigma\alpha\beta}+f_2\,\sqrt{g}\, \varepsilon^{\mu\nu\rho\sigma}F_{\mu\nu}F_{\rho \sigma}.
\end{equation}
Notice that $\mathcal{\tilde{A}}$ differs from the anomaly $\mathcal{A}$ defined in the previous sections by a factor $\sqrt{g}$.
Since we are using the Levi-Civita pseudotensor $\varepsilon^{\mu\nu\rho\sigma}$ with $\varepsilon^{0123}=\frac{1}{\sqrt{g}}$, the $\sqrt{g}$ dependence in \eqref{eq:atildeaa} cancels out. Therefore, the gauge term $F\tilde{F}$ in $\mathcal{\tilde{A}}$ is metric independent. This point will be crucial in our following analysis. 

\subsection{The general structure of the anomalous CWIs}
Since we are not invoking any renormalization of the correlators but only the presence of a chirally odd trace anomaly at $d=4$, we need to trace back the steps of a derivation of the anomalous CWIs at $d=4$, derived in \cite{Coriano:2017mux}. 
For the derivation of the dilatation and the special CWIs we use the conservation of the conformal currents
\be
J^{\m}_{(K)}(x) \equiv K_\n(x)\, T^{\m\n}(x)
\label{Jxi}
\ee
expressed in terms of the $K_{\m}(x)$s, which are the Conformal Killing Vectors (CKVs) satisfying the equation 
\be
\pa_{(\m}K_{\n)} \equiv \sdfrac{1}{2} \,\big(\pa_{\m}K_{\n} + \pa_{\n} K_{\m}\big) = \sdfrac{1}{d} \, \eta_{\m\n} \,\left(\pa\cdot K\right)
\label{CKeq}
\ee
in a $d$ dimensional Minkowski space. The conservation of $J^{\m}_{(K)}$ is violated in the case of a conformal anomaly. The current (\ref{Jxi}) can be inserted in the $n$-point stress tensor correlator to derive anomalous CWIs. Thus, we are led to consider total divergences of the form
\bea
&&\hspace{-1.2cm}\frac{\pa}{\pa x^\n} \Big\{ K_{\m}(x)\, \cS_{n+1}^{\m\n\m_1\n_1\dots\m_n\n_n}(x, x_1,\dots, x_n) \Big\}\nn \\
&&\hspace{-1cm} =K_{\m}(x)\,\pa_{\n}  \cS_{n+1}^{\m\n\m_1\n_1\dots\m_n\n_n}(x, x_1,\dots, x_n) 
 + \frac{1}{d} \, \left(\pa\cdot K\right)\,\eta_{\m\n} \,\cS_{n+1}^{\m\n\m_1\n_1\dots\m_n\n_n}(x, x_1,\dots, x_n),
 \label{CWIgen}
\eea
 where (\ref{CKeq}) has been used, in order to derive the CWIs for the $n$-point functions of the stress tensor.  The derivation of the identities require an integration of the previous equation over the spacetime points, assuming that we can drop a boundary term. In this case, since the trace anomaly introduces a term that can, in principle, be supported at spacetime infinity, the integration needs to be checked a posteriori. \\
The derivation of this equation can be found in \cite{Coriano:2017mux} for the $\langle TTT\rangle $, while the derivation for the 
$\langle JJT\rangle $ will be worked out below.
If we use on the rhs of \eqref{CWIgen} the conservation and anomalous trace Ward Identities for the $(n+1)$-point function $\cS_{n+1}$, we can derive the dilatation CWIs for $K$ of the form 
\be
\label{CK1}
K_{\m}^{(D)}(x) \equiv  x_\m\,,\qquad\qquad \pa\cdot K^{(D)} = d,
\ee
whereas the special CWIs are derived  by introducing $d$ Special Conformal Killing vectors in flat space
\be
\label{CK2}
K^{(C)\, \ka}_{\ \ \ \m}(x) \equiv 2x^{\ka}x_{\m} - x^2 \del^{\ka}_{\ \,\m}\,,
\qquad\qquad \pa\cdot K^{(C)\,\ka}(x) = (2d)\,x^{\ka}\,,\quad\ka =1, \dots, d 
\ee
which generate all the CWIs when these 4-vectors are substituted into (\ref{CWIgen}). For $n=3$, denoting the conformal weights of the operators with $\Delta_i$, we have
\begin{equation}
	\sum_{i=1}^3\left(\Delta_i+x_i^\mu \frac{\partial}{\partial x_i^\mu}\right)\left\langle T^{\mu_1\nu_1}\left(x_1\right) T^{\mu_2\nu_2}\left(x_2\right) T^{\mu_3 \nu_3}\left(x_3\right)\right\rangle=2^3 \int d x\,  \frac{\delta^3 \tilde{\mathcal{A}}(x)}{\delta g_{\mu_1\nu_1}\left(x_1\right) \delta g_{\mu_2\nu_2}\left(x_2\right) \delta g_{\mu_3 \nu_3}\left(x_3\right)}
\label{DTTT}
\end{equation}
for the dilatation  WI, and 
\small
\begin{align}
	&\sum_{i=1}^3\left[2 x_i^\kappa\left(\Delta_i+x_i^\alpha \frac{\partial}{\partial x_i^\alpha}\right)-x_i^2 \delta^{\kappa \alpha} \frac{\partial}{\partial x_i^\alpha}\right]
	\braket{T^{\mu_1\nu_1}(x_1)T^{\mu_2\nu_2}(x_2)T^{\mu_3\nu_3}({x}_3)}\nonumber\\&+
	2\bigg[\delta^{\kappa \mu_1}{x_1}_\alpha-\delta^\kappa_\alpha x_1^{\mu_1}
	\bigg]
	\braket{T^{\alpha\nu_1}(x_1)T^{\mu_2\nu_2}(x_2)T^{\mu_3\nu_3}({x}_3)}
	+
	2\bigg[\delta^{\kappa \nu_1}{x_1}_\alpha-\delta^\kappa_\alpha x_1^{\nu_1}
	\bigg]
	\braket{T^{\mu_1\alpha}(x_1)T^{\mu_2\nu_2}(x_2)T^{\mu_3\nu_3}({x}_3)}
	\nonumber\\&+
	2\bigg[\delta^{\kappa \mu_2}{x_2}_\alpha-\delta^\kappa_\alpha x_2^{\mu_2}
	\bigg]
	\braket{T^{\mu_1\nu_1}(x_1)T^{\alpha\nu_2}(x_2)T^{\mu_3\nu_3}({x}_3)}
	+
	2\bigg[\delta^{\kappa \nu_2}{x_2}_\alpha-\delta^\kappa_\alpha x_2^{\nu_2}
	\bigg]
	\braket{T^{\mu_1\nu_1}(x_1)T^{\mu_2\alpha}(x_2)T^{\mu_3\nu_3}({x}_3)}
	\nonumber\\&+
	2\bigg[\delta^{\kappa \mu_3}{x_3}_\alpha-\delta^\kappa_\alpha x_3^{\mu_3}
	\bigg]
	\braket{T^{\mu_1\nu_1}(x_1)T^{\mu_2\nu_2}(x_2)T^{\alpha\nu_3}({x}_3)}
	+
	2\bigg[\delta^{\kappa \nu_3}{x_3}_\alpha-\delta^\kappa_\alpha x_3^{\nu_3}
	\bigg]
	\braket{T^{\mu_1\nu_1}(x_1)T^{\mu_2\nu_2}(x_2)T^{\mu_3\alpha}({x}_3)}
	\nonumber\\&\hspace{5cm}=
	2^4 \int dx \, x^\kappa \frac{\delta^3 \tilde{ \mathcal{A}}(x) }{\delta g_{\mu_1\nu_1}(x_1) \delta g_{\mu_2\nu_2}(x_2)\delta g_{\mu_3\nu_3}(x_3)} 
\label{STTT}
\end{align}
\normalsize
for the special conformal WIs. We are going to discuss  the structure of 
these equations in the following, but first we turn to the analogous equations for the $\langle JJT\rangle$, $\langle JJO\rangle$ and $\langle TTO\rangle $ correlators. 
\subsection{The $\langle JJT\rangle$}
A similar analysis can be done for the $\langle JJT\rangle$, as we are now going to show, since it has not been given before.\\
We start by assuming that the following surface terms vanish, due to the fast fall-off behaviour of the correlation function
at infinity
\begin{equation}
	\begin{aligned}
		0=\int& dx \, \partial_\mu \bigg[K_\nu \langle T^{\mu\nu}(x)J^{\mu_1}(x_1)J^{\mu_2}(x_2)T^{\mu_3\nu_3}(x_3)\rangle \bigg]=\\&
		\int dx \,\left( \partial_\mu K_\nu\right) \langle 	T^{\mu\nu}(x)J^{\mu_1}(x_1)J^{\mu_2}(x_2)T^{\mu_3\nu_3}(x_3)\rangle+
		K_\nu \partial_\mu \langle T^{\mu\nu}(x)J^{\mu_1}(x_1)J^{\mu_2}(x_2)T^{\mu_3\nu_3}(x_3)\rangle.
	\end{aligned}
\end{equation}
Recalling the conformal Killing vector equation \eqref{CKeq}, we can then write
\begin{equation}\label{eq:ckvtjjt}
	\begin{aligned}
		0=
			 \int dx \,\left( \frac{\partial\cdot K}{d}\right)\eta_{\mu\nu} \langle T^{\mu\nu}(x)J^{\mu_1}(x_1)J^{\mu_2}(x_2)T^{\mu_3\nu_3}(x_3)\rangle+
			 K_\nu \partial_\mu \langle T^{\mu\nu}(x)J^{\mu_1}(x_1)J^{\mu_2}(x_2)T^{\mu_3\nu_3}(x_3)\rangle.
	\end{aligned}
\end{equation}
On the right-hand of the last equation we have the trace and the divergence of a four-point correlator function. We can use the anomalous trace equation and the conservation of the energy-momentum tensor in order to rewrite such terms. We will show this in the following.\\
We first focus on the dilatations. The Killing vectors in this case are given by \eqref{CK1}.  \\
The invariance under diffeomorphism leads to
\begin{equation}
	\nabla_\mu \langle T^{\mu \nu}\rangle-F^{\nu\mu}\langle J_\mu\rangle+A^\nu \, \nabla\cdot\langle J \rangle=0.
\end{equation}
Applying functional derivatives to this last equation and going to the flat limit, we obtain
\small
\begin{equation}\label{eq:divtjjt}
	\begin{aligned}
		0=&\partial_\mu \langle T^{\mu\nu}(x)J^{\mu_1}(x_1)J^{\mu_2}(x_2)T^{\mu_3\nu_3}(x_3)\rangle-\eta^{\mu_3\nu_3} \left(\partial_\mu\delta_{xx_3}\right) \langle J^{\mu_1}(x_1)J^{\mu_2}(x_2)T^{\mu\nu}(x)\rangle+\\&
		\bigg[-\delta^{\mu_1}_\mu \partial_\nu \delta_{xx_1} + \delta^{\mu_1}_\nu \partial_\mu \delta_{xx_1}\bigg]\langle J^{\mu}(x)J^{\mu_2}(x_2)T^{\mu_3\nu_3}(x_3)\rangle+
		\bigg[-\delta^{\mu_2}_\mu \partial_\nu \delta_{xx_2} + \delta^{\mu_2}_\nu \partial_\mu \delta_{xx_2}\bigg]\langle J^{\mu_1}(x_1)J^{\mu}(x)T^{\mu_3\nu_3}(x_3)\rangle\\&+
		\frac{1}{2}\langle J^{\mu_1}(x_1)J^{\mu_2}(x_2)T^{\mu \lambda}(x)\rangle\bigg[\delta_\lambda^{\nu_3}\delta^{\nu\mu_3}\partial_\mu \delta_{xx_3}+\delta_\mu^{\nu_3}\delta^{\nu\mu_3}\partial_\lambda\delta_{xx_3}-\delta_\lambda^{\nu_3}\delta_\mu^{\mu_3}\partial^\nu\delta_{xx_3}+\\&
		\delta_\lambda^{\mu_3}\delta^{\nu\nu_3}\partial_\mu \delta_{xx_3}+\delta_\mu^{\mu_3}\delta^{\nu\nu_3}\partial_\lambda\delta_{xx_3}-\delta_\lambda^{\mu_3}\delta_\mu^{\nu_3}\partial^\nu\delta_{xx_3}
		\bigg]+\delta^{\mu_3\nu_3}\left(\partial_\lambda\delta_{xx_3}\right)\langle J^{\mu_1}(x_1)J^{\mu_2}(x_2)T^{\lambda\nu}(x)\rangle,
	\end{aligned}
\end{equation}
\normalsize
where we denoted with $\delta_{xy}$ the Dirac delta function $\delta^4(x-y)$.
The anomalous trace equation is instead given by
\begin{equation} 
	0=2 g_{\mu \nu}(x) \frac{\delta \mathcal{S}[g]}{\delta g_{\mu \nu}(x)}-\tilde{\mathcal{A}}(x).
\end{equation}
Applying once again functional derivatives to this last equation and going to the flat limit, we have
\begin{equation}\label{eq:trtjjt}
	\begin{aligned}
		0=\eta_{\mu\nu}&\langle T^{\mu\nu}(x)J^{\mu_1}(x_1)J^{\mu_2}(x_2)T^{\mu_3\nu_3}(x_3)\rangle\\&
		+2\delta_{xx_3}\langle J^{\mu_1}(x_1)J^{\mu_2}(x_2)T^{\mu_3\nu_3}(x_3)\rangle-2\frac{\delta^3 \tilde{ \mathcal{A}}(x) }{\delta A_{\mu_1}(x_1) \delta A_{\mu_2}(x_2)\delta g_{\mu_3\nu_3}(x_3)} .
	\end{aligned}
\end{equation}

Inserting the equations \eqref{CK1}, \eqref{eq:divtjjt} and \eqref{eq:trtjjt} into \eqref{eq:ckvtjjt} and integrating by parts we arrive to
\begin{equation}
	\sum_{i=1}^3 \left( \Delta_i + x_i^\mu \frac{\partial}{\partial x_i^\mu}\right)	\langle J^{\mu_1}(x_1)J^{\mu_2}(x_2)T^{\mu\nu} ({x}_3)\rangle=2\int dx \frac{\delta^3 \tilde{ \mathcal{A}}(x) }{\delta A_{\mu_1}(x_1) \delta A_{\mu_2}(x_2)\delta g_{\mu_3\nu_3}(x_3)} 
\label{DJJT}
\end{equation}
for the dilatation CWI. \\
If, instead ,we consider the special conformal transformations, the Killing vectors are given in \eqref{CK2}.
Proceeding in a similar manner we arrive at the expression

  \begin{equation}
	\begin{aligned}
		&\sum_{i=1}^3\bigg[2 x_i^\kappa  \left(\Delta_i+x_i^\alpha \frac{\partial}{\partial x_i^\alpha}\right)-x_i^2 \delta^{\kappa \alpha} \frac{\partial}{\partial x_i^\alpha}\bigg]\left\langle J^{\mu_1}\left(x_1\right) J^{\mu_2}\left(x_2\right) T^{\mu_3 \nu_3}\left(x_3\right)\right\rangle \\&+
		2\bigg[\delta^{\kappa \mu_1} x_{1 \alpha}-\delta_\alpha^\kappa x_1^{\mu_1}\bigg]\left\langle J^\alpha\left(x_1\right) J^{\mu_2}\left(x_2\right) T^{\mu_3 \nu_3}\left(x_3\right)\right\rangle+2\bigg[\delta^{\kappa \mu_2} x_{2 \alpha}-\delta_\alpha^\kappa x_2^{\mu_2}\bigg]\left\langle J^{\mu_1}\left(x_1\right) J^\alpha\left(x_2\right) T^{\mu_3 \nu_3}\left(x_3\right)\right\rangle
		\\&+
		2\bigg[\delta^{\kappa \mu_3}{x_3}_\alpha-\delta^\kappa_\alpha x_3^{\mu_3}
		\bigg]
		\braket{J^{\mu_1}(x_1)J^{\mu_2}(x_2)T^{\alpha\nu_3}({x}_3)}
		+
		2\bigg[\delta^{\kappa \nu_3}{x_3}_\alpha-\delta^\kappa_\alpha x_3^{\nu_3}
		\bigg]
		\braket{J^{\mu_1}(x_1)J^{\mu_2}(x_2)T^{\mu_3\alpha}({x}_3)}
		 \\&\hspace{5cm}
		=2^2 \int d x x^\kappa \frac{\delta^3 \tilde{\mathcal{A}}(x)}{\delta A_{\mu_1}\left(x_1\right) \delta A_{\mu_2}\left(x_2\right) \delta g_{\mu_3 \nu_3}\left(x_3\right)}
	\end{aligned}
\label{SJJT}
\end{equation}
for the special CWI.

\subsection{$\langle JJO\rangle$  and $\langle TTO\rangle$}
The analysis of the subtraction terms in the anomaly \eqref{eq:defanomduff} will be taken into account by the inclusion of a scalar or pseudoscalar operator, indicated with $O$. In the case of the $\langle JJO\rangle $ we obtain for the dilatation WI
\begin{equation}
	\sum_{i=1}^3\left(\Delta_i+x_i^\mu \frac{\partial}{\partial x_i^\mu}\right)\left\langle J^{\mu_1}\left(x_1\right) J^{\mu_2}\left(x_2\right) O \left(x_3\right)\right\rangle= \int d x \frac{\delta^3 \tilde{\mathcal{A}}(x)}{\delta A_{\mu_1}\left(x_1\right) \delta A_{\mu_2}\left(x_2\right) \delta \phi_0 \left(x_3\right)}
\label{DJJO}
\end{equation}
and
\begin{align}
	\sum_{i=1}^3\bigg[2 x_i^\kappa& \left(\Delta_i+x_i^\alpha \frac{\partial}{\partial x_i^\alpha}\right)-x_i^2 \delta^{\kappa \alpha} \frac{\partial}{\partial x_i^\alpha}\bigg]
	\braket{J^{\mu_1}(x_1)J^{\mu_2}(x_2)O({x}_3)}+\nonumber\\&
	2\bigg[\delta^{\kappa \mu_1}{x_1}_\alpha-\delta^\kappa_\alpha x_1^{\mu_1}
	\bigg]
	\braket{J^{\alpha}(x_1)J^{\mu_2}(x_2)O({x}_3)}+
	2\bigg[\delta^{\kappa \mu_2}{x_2}_\alpha-\delta^\kappa_\alpha x_2^{\mu_2}
	\bigg]
	\braket{J^{\mu_1}(x_1)J^{\alpha}(x_2)O({x}_3)}
	\nonumber\\&\hspace{5cm}=
	2 \int dx \, x^\kappa \frac{\delta^3 \tilde{ \mathcal{A}}(x) }{\delta A_{\mu_1}(x_1) \delta A_{\mu_2}(x_2)\delta \phi_0(x_3)} 
\label{SJJO}
\end{align}
for the special CWI, while for the $\langle TTO\rangle $ the dilatation WI is
\begin{equation}
	\sum_{i=1}^3\left(\Delta_i+x_i^\mu \frac{\partial}{\partial x_i^\mu}\right)\left\langle T^{\mu_1\nu_1}\left(x_1\right) T^{\mu_2\nu_2}\left(x_2\right) O\left(x_3\right)\right\rangle=2^2 \int d x\,  \frac{\delta^3 \tilde{\mathcal{A}}(x)}{\delta g_{\mu_1\nu_1}\left(x_1\right) \delta g_{\mu_2\nu_2}\left(x_2\right) \delta \phi_0\left(x_3\right)},
	\label{DTTO}
\end{equation}
and the special CWI is 
\begin{align}
	&\sum_{i=1}^3\left[2 x_i^\kappa\left(\Delta_i+x_i^\alpha \frac{\partial}{\partial x_i^\alpha}\right)-x_i^2 \delta^{\kappa \alpha} \frac{\partial}{\partial x_i^\alpha}\right]
	\braket{T^{\mu_1\nu_1}(x_1)T^{\mu_2\nu_2}(x_2)O({x}_3)}\nonumber\\&+
	2\bigg[\delta^{\kappa \mu_1}{x_1}_\alpha-\delta^\kappa_\alpha x_1^{\mu_1}
	\bigg]
	\braket{T^{\alpha\nu_1}(x_1)T^{\mu_2\nu_2}(x_2)O({x}_3)}
	+
	2\bigg[\delta^{\kappa \nu_1}{x_1}_\alpha-\delta^\kappa_\alpha x_1^{\nu_1}
	\bigg]
	\braket{T^{\mu_1\alpha}(x_1)T^{\mu_2\nu_2}(x_2)O({x}_3)}
	\nonumber\\&+
	2\bigg[\delta^{\kappa \mu_2}{x_2}_\alpha-\delta^\kappa_\alpha x_2^{\mu_2}
	\bigg]
	\braket{T^{\mu_1\nu_1}(x_1)T^{\alpha\nu_2}(x_2)O({x}_3)}
	+
	2\bigg[\delta^{\kappa \nu_2}{x_2}_\alpha-\delta^\kappa_\alpha x_2^{\nu_2}
	\bigg]
	\braket{T^{\mu_1\nu_1}(x_1)T^{\mu_2\alpha}(x_2)O({x}_3)}
	\nonumber\\&\hspace{5cm}=
	2^3 \int dx \, x^\kappa \frac{\delta^3 \tilde{ \mathcal{A}}(x) }{\delta g_{\mu_1\nu_1}(x_1) \delta g_{\mu_2\nu_2}(x_2)\delta \phi_0(x_3)}. 
\label{STTO}
\end{align}
In the equations above we have introduced a coupling of the operator $O$ to an external source $\phi_0(x)$.

\subsection{The $\langle TTT\rangle$ in momentum space} 
The structure of the anomalous CWIs in momentum space are defined by applying a Fourier transform. We recall their general structure in the case of the $\langle TTT\rangle$, the other being similar.\\
We first introduce the Fourier transform of the functional derivative of the anomaly functional in the form 
\bea
&&\hspace{-1.5cm} (2\pi)^4 \,\d^4 (p_1+ \dots + p_{n+1})\, \tilde \cA_{n}^{\m_2\n_2\dots\m_{n+1}\n_{n+1}}(p_1,\dots ,p_{n+1}) \nn\\
&& \quad \equiv \int d^4x_1\dots d^4x_{n+1} \ e^{i p_1\cdot x_1 + \dots + i p_{n+1}\cdot x_{n+1}} \, 
\frac{\del^{n}\tilde{\cA} (x_1)}{\del g_{\m_2\n_2}(x_2)\dots \del g_{\m_{n+1}\n_{n+1}}(x_{n+1})}\Big\vert_{flat}.
\label{Avardef}
\eea
We can then write the anomalous Dilatation CWI
in momentum space 
\begin{equation}
\left[4 - p_1\cdot\sdfrac{\pa}{\pa p_1} - p_2\cdot\sdfrac{\pa}{\pa p_2} \right]
\langle T^{\mu_1\nu_1}(p_1)T^{\mu_2\nu_2}(p_2)T^{\mu_3\nu_3}(-p_1-p_2)\rangle
 = 8 \, \tilde \cA_3^{\m_1\n_1\m_2\n_2\m_3\n_3}(p_1,p_2,-p_1-p_2),\label{CWIs3a}
\end{equation}
where we set $\Delta_i=d=4$. The special CWI are instead given by
\bea
&&\hspace{-1cm} \sum_{j=1}^2 \left[ -2\, p_{j \a}  \frac{\pa^2}{\pa p_{j\a} \pa p_{j  \ka}} + p_j^{\ka} \frac{\pa^2}{\pa p_{j\a} \pa p_j^\a} \right]
\langle T^{\mu_1\nu_1}(p_1)T^{\mu_2\nu_2}(p_2)T^{\mu_3\nu_3}(-p_1-p_2)\rangle \nn \\
&& +\,  2 \left(\h^{\ka\m_1} \frac{\pa}{\pa p_1^{\a_1}} - \d^{\ka}_{\ \a_1} \frac{\pa}{\pa p_{1\m_1}}\right) \langle T^{\mu_1\nu_1}(p_1)T^{\mu_2\nu_2}(p_2)T^{\mu_3\nu_3}(-p_1-p_2)\rangle\nn \\
&& +\,  2 \left(\h^{\ka\n_1} \frac{\pa}{\pa p_1^{\b_1}} - \d^{\ka}_{\ \b_1} \frac{\pa}{\pa p_{1\n_1}}\right) \langle T^{\mu_1\nu_1}(p_1)T^{\mu_2\nu_2}(p_2)T^{\mu_3\nu_3}(-p_1-p_2)\rangle\nn \\
&& +\,  2 \left(\h^{\ka\m_2} \frac{\pa}{\pa p_2^{\a_2}} - \d^{\ka}_{\ \a_2} \frac{\pa}{\pa p_{2\m_2}}\right) \langle T^{\mu_1\nu_1}(p_1)T^{\mu_2\nu_2}(p_2)T^{\mu_3\nu_3}(-p_1-p_2)\rangle\nn \\
&& +\,  2 \left(\h^{\ka\n_2} \frac{\pa}{\pa p_2^{\b_2}} - \d^{\ka}_{\ \n_2} \frac{\pa}{\pa p_{2\n_2}}\right) \langle T^{\mu_1\nu_1}(p_1)T^{\mu_2\nu_2}(p_2)T^{\mu_3\nu_3}(-p_1-p_2)\rangle\nn \\
&&\hspace{1cm} = -16\ \frac{\pa}{\pa p_{3\ka}} \tilde \cA_3^{\m_1\n_1\m_2\n_2\m_3\n_3}(p_1,p_2,p_3)\Big\vert_{p_3 = -p_1 -p_2}.\label{CWIs3b}
\eea
 Notice that the presence of a differentiation of $\mathcal{A}_3$ with respect to one of the momenta on the rhs of the last equation \eqref{CWIs3b} is what makes the anomalous term nonzero in the case of a topological anomaly. The differentiation comes from the presence of the factor $x^{\kappa}$ in the integration of \eqref{STTT}. The anomalous contribution is present, naturally, in all the CWIs, but in some cases it can vanishes, as we are going to show below.

\subsection{The non anomalous character of the CWIs}
Let us have a closer look at the rhs of conformal equations, in the case in which the anomaly contains topological terms of the form $R\tilde R$ 
and $F\tilde F$. We are going to show that for some of these equations the anomalous contribution vanishes. In other words, they correspond to ordinary (i.e. non-anomalous) CWIs.  \\
In the case of a topological anomaly, the term on the right-hand sides of \eqref{DTTT},\eqref{DJJT},{\eqref{DJJO},\eqref{DTTO}, always vanish.
Indeed, topological anomalies are scale-invariant: they do not break dilatations.
This can be seen for example in the case of the dilatation equations of the $\langle TTT \rangle$ in \eqref{DTTT}.
If we commute the integration and the functional differentiation with respect to the metric 
on the rhs of the equations we get a vanishing result
\begin{equation}
	\int d x\,  \frac{\delta^3 \tilde{\mathcal{A}}(x)}{\delta g_{\mu_1\nu_1}\left(x_1\right) \delta g_{\mu_2\nu_2}\left(x_2\right) \delta g_{\mu_3 \nu_3}\left(x_3\right)}=\,  \frac{\delta^3}{\delta g_{\mu_1\nu_1}\left(x_1\right) \delta g_{\mu_2\nu_2}\left(x_2\right) \delta g_{\mu_3 \nu_3}\left(x_3\right)}  \int d x\tilde{\mathcal{A}(x)}=0. 
\end{equation}
However special conformal transformations can potentially be broken by a topological anomaly. Indeed, the anomalous term on the right-hand side of the equations \eqref{STTT}, \eqref{SJJT}, \eqref{SJJO},  \eqref{STTO}  in principle can be non-vanishing due to the presence of $x^\kappa$ in the integrand, as mentioned above.
Let's illustrate this point case by case.\\
In the case of $\tilde{\mathcal{A}}(x)=  \sqrt{g} \, f \, \varepsilon^{\mu\nu\rho\sigma} F_{\mu\nu}F_{\rho\sigma}$, it is clear that the anomaly does not depend on the metric  since 
\begin{equation}
	\varepsilon^{\mu\nu\rho\sigma}=\frac{\epsilon^{\mu\nu\rho\sigma}}{\sqrt{g}}
\end{equation}
with $\epsilon^{0123}=1$.
Therefore, when applying a functional derivative with respect to the metric to $\tilde{\mathcal{A}}(x)$, we get a vanishing result
 \begin{equation}
 	 \frac{\delta^3 \tilde{ \mathcal{A}}(x) }{\delta A_{\mu_1}(x_1) \delta A_{\mu_2}(x_2)\delta g_{\mu_3\nu_3}(x_3)} =0.
\end{equation}
This shows that \eqref{DJJT} and \eqref{SJJT} are homogeneous and henceforth they are ordinary CWIs. A similar result is obtained in the case of \eqref{DJJO} and \eqref{SJJO} since
\begin{equation}
	\frac{\delta}{\delta \phi_0}=2 g_{\mu\nu}\frac{\delta}{\delta g_{\mu\nu}}.
\end{equation}  
Therefore, dilatations and special conformal transformations are ordinary in the case of the $\langle JJT\rangle$ and $\langle JJO\rangle$.\\
Let us now look at the conformal equations of the $\langle TTO \rangle $. The anomalous term is given by $R\tilde{R}$ which is Weyl invariant. Therefore, when acting on it with the operator $\frac{\delta}{\delta \phi_0}=2 g_{\mu\nu}\frac{\delta}{\delta g_{\mu\nu}}$ we get a vanishing result. As a consequence, once again the anomalous term vanishes in the equation \eqref{DTTO} and \eqref{STTO} and the correlator satisfies ordinary CWIs.
\\
The last correlator we need to consider is the $\langle TTT \rangle$. As we already said, dilatations are not broken by a topological anomaly. However one can check that in the case of the $\langle TTT \rangle$ the anomalous term does not vanish in the special conformal Ward identities.\\
We conclude that almost all of  all the CWIs we considered are ordinary and non-anomalous. The only anomalous 
CWI is the \eqref{STTT}, which is allowed to be nonzero in the presence of an $R\tilde{R}$ term in the anomaly functional.  

\section{The ${\langle JJO\rangle}_{odd}$ correlator in CFT}
In this section we study the conformal constraints on the $\braket{JJO}$ correlator where each $J$ can be a conserved vector current or an anomalous axial-vector current.
Our result is valid for every odd parity $\braket{JJO}$ correlator constructed with two potentially different (axial and/or conserved) currents and a scalar/pseudoscalar operator.
As we will see, the solution of the CWIs of the correlator can be written in terms of 3K integrals that needs a regularization. 
Therefore, we work in the general scheme $\{ u,v\}$ where $d=4+2 u \epsilon $ and the conformal dimensions $\Delta_i$ of the operators is shifted by $(u+v_i) \, \epsilon $ with arbitrary $u$ and $v_i$.\\
We start by decomposing the operators $J$ in terms of their transverse part and longitudinal ones (also termed "local") \cite{Bzowski:2013sza}
\begin{align}
	J^{\mu_i}(p_i)&\equiv j^{\mu_i}(p_i)+j_{loc}^{\mu_i}(p_i),
\end{align}
where
\begin{align}
	j^{\mu_i}(p_i)&=\pi^{\mu_i}_{\alpha_i}(p_i)\,J^{\alpha_i }(p_i), &&\hspace{1ex}j_{loc}^{\mu_i}(p_i)=\frac{p_i^{\mu_i}\,p_{i\,\alpha_i}}{p_i^2}\,J^{\alpha_i}(p_i),
\end{align}
having introduced the transverse transverse projector 
\begin{align}
	\pi^{\mu}_{\alpha} & = \delta^{\mu}_{\alpha} - \frac{p^{\mu} p_{\alpha}}{p^2} .
\end{align}
We then consider the following conservation Ward identities
\begin{equation}
	\begin{aligned} \label{eq:CWIJJO}
		\nabla_\mu\langle {J^\mu_c}\rangle =0, \qquad\qquad \nabla_\mu \langle {J^\mu_5}\rangle=a \, 	\varepsilon^{\mu\nu\rho\sigma}F_{\mu\nu}F_{\rho\sigma}
	\end{aligned}
\end{equation}
of the expectation value of the conserved $J_c^{\mu}$ and anomalous $J_5^{\mu}$ currents.\\ The vector current is coupled to the 
vector source $V_{\mu}$ and the axial-vector current to the source $A_\mu$. The operator $O$ in the $\langle JJO\rangle$ is coupled to a scalar field source $\phi$.
Applying multiple functional derivatives to \eqref{eq:CWIJJO} with respect to the sources, after a Fourier transform, we find the conservation Ward identities related to the entire correlator
\begin{equation}\label{eq:jjoconservedid}
	\begin{aligned}
		p_{i\mu_i} \langle J^{\mu_1}(p_1)J^{\mu_2}(p_2)O(p_3)\rangle=0.\quad \quad i=1,2
	\end{aligned}
\end{equation}
Such equation is satisfied independently of the fact that $J$'s are conserved vector or axial-vector currents.
Indeed, the chiral anomaly does not contribute to the $\langle JJO\rangle$.
Due to the identities in \eqref{eq:jjoconservedid}, the longitudinal part of the correlator vanishes. On the other hand, the transverse part can be formally expressed in terms of the following tensor structure 
\begin{equation} \label{eq:decompFinJJO}
	\left\langle J^{ \mu_1}\left(p_1\right) 
	J^{   \mu_2}\left(p_2\right) O	\left(p_3\right)\right\rangle=\left\langle j^{ \mu_1}\left(p_1\right) 
	j^{   \mu_2}\left(p_2\right) O	\left(p_3\right)\right\rangle  =\pi^{\mu_1}_{\alpha_1}\left(p_1\right)
	\pi^{\mu_2}_{\alpha_2} \left(p_2\right) \Bigl[  A(p_1,p_2,p_3)\varepsilon^{\alpha_1\alpha_2 p_1 p_2} \Bigr].
\end{equation}
where $\varepsilon^{\alpha_1\alpha_2 p_1 p_2}\equiv \varepsilon^{\alpha_1\alpha_2 \rho \sigma}p_{1\rho} p_{2\sigma}$.
Notice that in this case one can omit the projectors $\pi^{\mu_i}_{\alpha_i}\left(p_i\right)$ since they act as an identity on the tensorial structure in the brackets.

\subsection{Dilatation and Special Conformal Ward Identities}
We start analysing the conformal constraints on the form factor $A(p_1,p_2,p_3)$. The invariance of the correlator under dilatation is reflected in the equation
\begin{align}
	\left(\sum_{i=1}^3\Delta_i-2d-\sum_{i=1}^2\,p_i^\mu\frac{\partial}{\partial p_i^\mu}\right)\braket{J^{\mu_1}(p_1)J^{\mu_2}(p_2)O (p_3)}=0.\label{DilttJJO}
\end{align}
By using the chain rule
\begin{align}
	\frac{\partial}{\partial p_i^\mu}=\sum_{j=1}^3\frac{\partial p_j}{\partial p_i^\mu}\frac{\partial}{\partial p_j},
\end{align}
in term of the invariants $p_i=|\sqrt{p_i^2}|$ and by considering the decomposition \eqref{eq:decompFinJJO},
 we can rewrite the dilatation equation as a constraint on the form factor $A$ 
\begin{align}
	\sum_{i=1}^{3} p_i \frac{\partial A}{\partial p_i }-\left(\sum_{i=1}^3\Delta_i-2d- N\right) A=0,
\end{align}
with $N=2$, the number of momenta that the form factor $A$ multiply in the decomposition.
\\
The invariance of the correlator with respect to the special conformal transformations is encoded in the special conformal Ward identities 
\begin{align}
	0=&\sum_{j=1}^2\left[-2\frac{\partial}{\partial p_{j\kappa}}-2p_j^\alpha\frac{\partial^2}{\partial p_j^\alpha\,\partial p_{j\kappa}}+p_j^\kappa\frac{\partial^2}{\partial p_j^\alpha\,\partial p_{j\alpha}}\right]\braket{J^{\mu_1}(p_1)J^{\mu_2}(p_2)O(p_3)}\notag\\
	&+2\left(\delta^{\mu_1\kappa}\,\frac{\partial}{\partial p_1^{\alpha_1}}-\delta^\kappa_{\alpha_1}\,\frac{\partial}{\partial p_{1\mu_1}}\right)\braket{J^{\alpha_1}(p_1)J^{\mu_2}(p_2)O(p_3)}\notag\\
	&+2\left(\delta^{\mu_2\kappa}\,\frac{\partial}{\partial p_2^{\alpha_2}}-\delta^\kappa_{\alpha_2}\,\frac{\partial}{\partial p_{2\mu_2}}\right)\braket{J^{\mu_1}(p_1)J^{\alpha_2}(p_2)O(p_3)}\equiv \mathcal{K}^\kappa\braket{J^{\mu_1}(p_1)J^{\mu_2}(p_2)O(p_3)}.
\end{align}
The special conformal operator $\mathcal{K}^\kappa$ acts as an endomorphism on the transverse sector of the entire correlator. 
We can then perform a transverse projection on all the indices in order to identify a set of partial differential equations  
\begin{align}
	&\pi_{\mu_1}^{\lambda_1}(p_1)
	\pi_{\mu_2}^{\lambda_2} (p_2) \bigg(\mathcal{K}^\kappa\braket{J^{\mu_1}(p_1)J^{\mu_2}(p_2)O(p_3)}\bigg)=0.
\end{align} 
We then decompose the action of the special conformal operator on the transverse sector in the following way
\begin{equation} \label{eq:decompKJJOnonmin}
	\begin{aligned}
		0=&\pi_{\mu_1}^{\lambda_1}\left(p_1\right)
		\pi_{\mu_2}^{\lambda_2} \left(p_2\right) 
		\mathcal{K}^k\left\langle J^{ \mu_1}\left(p_1\right) 
		J^{  \mu_2}\left(p_2\right) O	\left(p_3\right)\right\rangle =
		\pi_{\mu_1}^{\lambda_1}\left(p_1\right)
		\pi_{\mu_2}^{\lambda_2} \left(p_2\right) 
		\Bigl[\\ &
		C_{11} \epsilon^{p_1 p_2 \mu_1 \mu_2} p_1^\kappa +	{C}_{21} \epsilon^{p_1 p_2 \mu_1 \mu_2} p_2^\kappa+ 
		C_{31} \epsilon^{p_1 \kappa \mu_1 \mu_2 }+{C}_{32} \epsilon^{p_2 \kappa \mu_1 \mu_2 }+
		C_{33} p_2^{\mu_1}\epsilon^{p_1 p_2 \kappa\mu_2 }+
		{C}_{34} p_3^{\mu_2}\epsilon^{p_1 p_2 \kappa\mu_1 }\Bigr],
	\end{aligned}
\end{equation}
where $C_{ij}$ are scalar functions of the form factor $A$ and its derivatives.
The tensor structures we have written are not all independent and can be simplified in order to find the minimal decomposition, using the following Schouten identities
\begin{equation}
	\begin{aligned}
		&\epsilon^{[p_1 p_2 \mu_1 \mu_2} p_1^{\kappa]}=0,\\
		&\epsilon^{[p_1 p_2 \mu_1 \mu_2} p_2^{\kappa]}=0,
	\end{aligned}
\end{equation}
according to which we can eliminate $C_{33}$ and ${C}_{34}$ 
\small
\begin{equation} \label{eq:}
	\begin{aligned}
		\pi_{\mu_1}^{\lambda_1}(p_1)\pi_{\mu_2}^{\lambda_2} (p_2)\bigg(
		\epsilon^{p_1 p_2 {\kappa}{\mu_1}} p_3^{\mu_2}\bigg) &=
		\pi_{\mu_1}^{\lambda_1}(p_1)\pi_{\mu_2}^{\lambda_2} (p_2)\,
		\bigg(
		\frac{1}{2} \epsilon^{p_1{\kappa}{\mu_1}{\mu_2}} (p_1^2+p_2^2-p_3^2)+\epsilon^{p_1p_2{\mu_1}{\mu_2}} p_1^{\kappa}+\epsilon^{p_2{\kappa}{\mu_1}{\mu_2}} p_1^2\bigg)\\
		\pi_{\mu_1}^{\lambda_1}(p_1)\pi_{\mu_2}^{\lambda_2} (p_2)\bigg(
		\epsilon^{p_1p_2{\kappa}{\mu_2}} p_2^{\mu_1}\bigg)&=
		\pi_{\mu_1}^{\lambda_1}(p_1)\pi_{\mu_2}^{\lambda_2} (p_2)\, \bigg(
		-\frac{1}{2} \epsilon^{p_2{\kappa}{\mu_1}{\mu_2}} (p_1^2+p_2^2-p_3^2)
		+\epsilon^{p_1p_2{\mu_1}{\mu_2}} p_2^{\kappa}
		-\epsilon^{p_1{\kappa}{\mu_1}{\mu_2}} p_2^2\bigg).
	\end{aligned}
\end{equation}
\normalsize
Therefore we can rewrite eq. \eqref{eq:decompKJJOnonmin} in the minimal form
\begin{equation}
	\begin{aligned}
		0=\pi_{\mu_1}^{\lambda_1}&\left(p_1\right)
		\pi_{\mu_2}^{\lambda_2} \left(p_2\right) 
		\mathcal{K}^k\left\langle J^{ \mu_1}\left(p_1\right) 
		J^{  \mu_2}\left(p_2\right) O	\left(p_3\right)\right\rangle =\\ &
		\pi_{\mu_1}^{\lambda_1}\left(p_1\right)
		\pi_{\mu_2}^{\lambda_2} \left(p_2\right) 
		\Bigl[
		C_{11} \epsilon^{p_1 p_2 \mu_1 \mu_2} p_1^\kappa +	{C}_{21} \epsilon^{p_1 p_2 \mu_1 \mu_2} p_2^\kappa+ 
		C_{31} \epsilon^{p_1 \kappa \mu_1 \mu_2 }+{C}_{32} \epsilon^{p_2 \kappa \mu_1 \mu_2 }\Bigr],	\end{aligned}
\end{equation}
where we have redefined the form factors $C_{ij}$ in order to include the contribution of the old $C_{33}$ and ${C}_{34}$.
Due to the independence of the tensor structures listed above, the special conformal equations can be written as
\begin{align}
	C_{ij}=0.
\end{align}
In particular, $C_{11}=0$ and $C_{21}=0$ are equations of the second order and therefore they are called primary equations \cite{Bzowski:2013sza}. All the others are first order differential equations and are called secondary equations.
The explicit form of the primary equations is
\begin{equation}
	\begin{aligned} \label{eq:primOmog} 
		K_{31}A=0,\\
		K_{32}A=0,
	\end{aligned}
\end{equation}
where we have defined
\begin{align}
	K_i=\frac{\partial^2}{\partial p_i^2}+\frac{(d+1-2\Delta_i)}{p_i}\frac{\partial}{\partial p_i},\qquad K_{ij}=K_i-K_j.
\end{align}
The secondary equations are
\begin{equation}
	\begin{aligned}
		0=p_2\frac{\partial A }{\partial p_2}+ (d-1-\Delta_2)A,\\
		0=p_1\frac{\partial A }{\partial p_1}+ (d-1-\Delta_1)A.
	\end{aligned}
\end{equation}
\subsection{Solution of the CWIs}
The most general solution of the conformal Ward identities of the $\langle JJO \rangle$ can be written in terms of integrals involving a product of three Bessel functions, namely 3K integrals. For a review on the properties of such integrals, see appendix \ref{appendix:3kint} and \cite{Bzowski:2013sza,Bzowski:2015pba,Bzowski:2015yxv}.
We recall the definition of the general 3K integral
\begin{equation}
	I_{\alpha\left\{\beta_1 \beta_2 \beta_3\right\}}\left(p_1, p_2, p_3\right)=\int d x x^\alpha \prod_{j=1}^3 p_j^{\beta_j} K_{\beta_j}\left(p_j x\right),
\end{equation}
where $K_\nu$ is a modified Bessel function of the second kind.
In particular, we will use the reduced version of the 3K integral defined as
\begin{equation}
	J_{N\left\{k_j\right\}}=I_{\frac{d}{2}-1+N\left\{\Delta_j-\frac{d}{2}+k_j\right\}},
\end{equation}
where we introduced the condensed notation $\{k_j \} = \{k_1, k_2, k_3 \}$.
The 3K integral satisfies an equation analogous to the dilatation equation with scaling degree
\begin{equation}
	\text{deg}\left(J_{N\left\{k_j\right\}}\right)=\Delta_t+k_t-2 d-N,
\end{equation}
where 
\begin{equation}
	k_t=k_1+k_2+k_3,\qquad\qquad \Delta_t=\Delta_1+\Delta_2+\Delta_3.
\end{equation}
From this analysis, it is simple to relate the form factors to the 3K integrals. Indeed, the dilatation Ward identities tell us that the form factor $A$ needs to be written as a combination of integrals of the following type
\begin{equation}
	J_{N+k_t,\{k_1,k_2,k_3\}},
\end{equation}
with $N=2$, the number of momenta that the form factor multiplies in the decomposition \eqref{eq:decompFinJJO}. The special conformal Ward identities fix the remaining indices $k_1$, $k_2$ and $k_3$.
Indeed, recalling the following property of the 3K integrals
\begin{equation}
	K_{n m} J_{N\left\{k_j\right\}}=-2 k_n J_{N+1\left\{k_j-\delta_{j n}\right\}}+2 k_m J_{N+1\left\{k_j-\delta_{j m}\right\}},
\end{equation}
we can write the most general solution of the primary equations \eqref{eq:primOmog} as
\begin{equation}
	A=c_1 J_{2\{0,0,0\}},
\end{equation} 
where $c_1$ is an arbitrary constant. Before moving on to the secondary equations, we need to discuss the possible divergences in 3K integrals which, in this case, can occur for some specific values of $\Delta_3$. In general, it can be shown that the 3K integral $I_{\alpha\{\beta_1,\beta_2,\beta_3\}}$ diverges if
\begin{equation}
	\alpha+1 \pm \beta_1 \pm \beta_2 \pm \beta_3=-2 k \quad, \quad k=0,1,2, \dots
\end{equation}
For a more detailed review of the topic, see appendix \ref{appendix:3kint} and \cite{Bzowski:2013sza,Bzowski:2015pba,Bzowski:2015yxv}.
If the above condition is satisfied, we need to regularize the integrals
\begin{equation}
	d \rightarrow 4+2 u \epsilon \quad  \quad \Delta_i \rightarrow \Delta_i+\left(u+v_i\right) \epsilon .
\end{equation}
In general, the regularisation parameters $u$ and $v_i$ are arbitrary. For simplicity, in this paper we will choose the same $v_i=v$ for each operator.
If a 3K integral in our solution diverges, we can expand the coefficient in front of such integral in the solution in powers of $\epsilon$ 
\begin{equation}
	c_i=  \sum_{j=-\infty}^\infty c_i^{(j)}\epsilon^j,
\end{equation}
and then we can require that our entire solution is finite for $\epsilon\rightarrow 0$ by constraining the coefficients $ c_i^{(j)}$.
Looking at our solution we can see that $J_{2\{0,0,0\}}\equiv I_{3+u\epsilon\{1+v \epsilon,1+v\epsilon,-2+\Delta_3 +v\epsilon \}}$ diverges for $\epsilon\rightarrow 0$ when
\begin{equation} \label{eq:dimconfcondiv}
	\Delta_3=0,4,6,8,10,\dots
\end{equation}
Let us now look at the secondary conformal equations
\begin{equation}
	\begin{aligned}
		0=p_2\frac{\partial A}{\partial p_2}-(\Delta_2-d+1)A\\
		0=p_1\frac{\partial A}{\partial p_1}-(\Delta_1-d+1)A.\\
	\end{aligned}
\end{equation}
We can solve such equation by performing the limit $p_i\rightarrow 0$ for various values of $\Delta_3$ (see appendix \ref{appendix:3kint} for a review of the procedure). If the eq. 
\eqref{eq:dimconfcondiv} is not satisfied and the 3K integral is finite, the secondary equations lead to the condition $c_1=\mathcal{O}(\epsilon)$ and therefore the entire correlator vanishes. However, in the case $\Delta_3=4$, the secondary equations still lead to $c_1=c_1^{(1)}\epsilon+\mathcal{O}(\epsilon^2)$ but since the following 3K integral has a pole \cite{Bzowski:2013sza,Bzowski:2015yxv,Bzowski:2020lip}
\begin{equation}
	I_{3+u\epsilon \{1+v\epsilon,1+v\epsilon,2+v\epsilon\}}=\frac{2}{(u-3v)\epsilon}+\mathcal{O}(\epsilon^0),
\end{equation}
the $\epsilon$ in the solution cancels out and we end up with a finite non-zero solution. 
The coefficient $c_1^{(1)}$ remains unconstrained. 
A similar story occurs when $\Delta_3=0$ since
\begin{equation}
	I_{3+u\epsilon\{1+v\epsilon,1+v\epsilon,-2+v\epsilon\}}=\frac{2}{(u-v) \, p_3^4 \, \epsilon }+\mathcal{O}(\epsilon^0).
\end{equation}
Note that such value of $\Delta_3$ is not physical and violates unitarity bounds. However, sometimes this may not constitute a problem if, for example, we are working in a holographic contest.
One can check that for $\Delta_3=6,8,10\dots$ and so on, the correlator vanishes. Indeed, in this cases the secondary equations require the coefficient $c_1$ to scale with high powers of $\epsilon$ that can not be compensated by the poles of the 3K integral $I_{3+u\epsilon\{1+v\epsilon,1+v\epsilon,-2+\Delta_3+v\epsilon\}}$. In the end we have
\begin{equation}
	\begin{aligned}
		&\left\langle J^{ \mu_1}\left(p_1\right) 
		J^{   \mu_2}\left(p_2\right) O_{(\Delta_3\neq {0,4})}	\left(p_3\right)\right\rangle =0\\
		&\left\langle J^{ \mu_1}\left(p_1\right) 
		J^{   \mu_2}\left(p_2\right) O_{(\Delta_3= 4)}	\left(p_3\right)\right\rangle =c^{(1)}_1 \varepsilon^{p_1 p_2 \mu_1\mu_2}\\
		&\left\langle J^{ \mu_1}\left(p_1\right) 
		J^{   \mu_2}\left(p_2\right) O_{(\Delta_3= 0)}	\left(p_3\right)\right\rangle =\frac{c^{(1)}_1}{p_3^4} \varepsilon^{p_1 p_2 \mu_1\mu_2},
	\end{aligned}
\end{equation}
where we have absorbed a factor $2/(u-3v)$ or $2/(u-v)$ in the constant $c^{(1)}_1$.
Therefore, excluding the unphysical case $\Delta_3=0$, the only other case where the correlator does not vanish is $\Delta_3=4$, which is satisfied for example if $O=\nabla_\mu J_5^\mu$ or $O=T^\mu_\mu$. Furthermore, it is important to note that the most general solution that we have found for the $\langle JJO\rangle$ with $\Delta_3=4$
can be written in terms of functional derivatives $F\tilde{F}$
\begin{equation}
	\begin{aligned}
		&\delta^4(p_1+p_2+p_3)\left\langle J^{ \mu_1}\left(p_1\right) 
		J^{   \mu_2}\left(p_2\right) O_{(\Delta_3= 4)}	\left(p_3\right)\right\rangle =  \\&
		\qquad\qquad \int  dx_1\, dx_2 \,dx_3\, e^{-i(p_1x_1+p_2x_2+p_3x_3)}\, \, \frac{\delta^2   \Big[f \, \varepsilon^{\mu \nu\rho\sigma}F_{\mu \nu}(x_3)F_{\rho\sigma}(x_3)\Big] }{\delta A_{\mu_1}(x_1)\, \, \delta A_{\mu_2} (x_2)},
	\end{aligned}
\end{equation}
in accord with the chiral anomaly formula \eqref{eq:anomaliachirale} for $O_{(\Delta_3= 4)}	=\nabla \cdot J_5$, and potentially a parity-odd trace anomaly $F\tilde{F}$ for the case $O_{(\Delta_3= 4)}	=T^\mu_\mu$.

\section{The $\langle JJT \rangle_{odd}$ correlator in CFT in the traceless case}
In section 3, we have found that the special and dilatation CWIs satisfied by the $\langle JJT \rangle$ are non-anomalous even in the presence of a trace anomaly $F\tilde{F}$. The natural question to ask is if a parity-odd trace anomaly is still allowed for these correlators, and what are the implications of these equations for the structure of their longitudinal and transverse components. The decomposition in such components has been introduced in \cite{Bzowski:2013sza} and discussed in several works, including their renormalization in a general CFT \cite{Bzowski:2015yxv,Bzowski:2018fql}. These former analysis, at least in the case of the $\langle TTT\rangle $ and $\langle TJJ\rangle$, in the parity even case, have been exactly matched in perturbation theory at one-loop and simplified, using three separate sectors of free field theories (Lagrangian) parameterizations. \\
The three constants appearing in the solution of the corresponding CWIs can be uniquely related at $d=4$ and $d=5$ with the multiplicities of scalars ($n_S$), fermions ($n_f$) and 
spin-1 gauge fields ($n_V$) of such free field theory realizations \cite{Coriano:2018bbe,Coriano:2018bsy}, providing a direct way to investigate the renormalization of the corresponding correlators using only ordinary Feynman diagrams. 
A general review of these methods can be found in \cite{Coriano:2020ees}. \\
As already mentioned in the previous sections, diagrams affected by topological anomalies do not need overall any regularization, the reason being that the possible counterterms that one could consider at one-loop in such realizations, are evanescent. However, it is well known from the analysis of the $\langle AVV\rangle$ interaction, where one faces such issues, that this diagram is affected by an ambiguity related to the momentum representation of the vertex, which can be avoided  by enforcing external Ward identities on the same vertex. \\
In this section we study the conformal constraints on the $\braket{JJT}_{odd}$ correlator. We assume that the correlator is not anomalous and therefore, if we trace over the energy-momentum, we get a vanishing result. We will relax such assumption in the next section. \\
We start the analysis by decomposing the operators $T$ and $J$ in terms of their transverse traceless part and longitudinal ones (also termed "local")
	\begin{align}
		T^{\mu_i\nu_i}(p_i)&= t^{\mu_i\nu_i}(p_i)+t_{loc}^{\mu_i\nu_i}(p_i),\label{decT}\\
		J^{\mu_i}(p_i)&= j^{\mu_i}(p_i)+j_{loc}^{\mu_i}(p_i),\label{decJ}
	\end{align}
	where
	\begin{align}
		\label{loct}
		t^{\mu_i\nu_i}(p_i)&=\Pi^{\mu_i\nu_i}_{\alpha_i\beta_i}(p_i)\,T^{\alpha_i \beta_i}(p_i), &&t_{loc}^{\mu_i\nu_i}(p_i)=\Sigma^{\mu_i\nu_i}_{\alpha_i\beta_i}(p)\,T^{\alpha_i \beta_i}(p_i),\\
		j^{\mu_i}(p_i)&=\pi^{\mu_i}_{\alpha_i}(p_i)\,J^{\alpha_i }(p_i), &&\hspace{1ex}j_{loc}^{\mu_i}(p_i)=\frac{p_i^{\mu_i}\,p_{i\,\alpha_i}}{p_i^2}\,J^{\alpha_i}(p_i),
	\end{align}
having introduced the transverse-traceless ($\Pi$), transverse $(\pi)$, longitudinal ($\Sigma$) projectors, given respectively by \small
\begin{equation}
	\label{prozero}
\begin{aligned}
	&\pi^{\mu}_{\alpha}  = \delta^{\mu}_{\alpha} - \frac{p^{\mu} p_{\alpha}}{p^2}, \\&
	\Pi^{\mu \nu}_{\alpha \beta}  = \frac{1}{2} \left( \pi^{\mu}_{\alpha} \pi^{\nu}_{\beta} + \pi^{\mu}_{\beta} \pi^{\nu}_{\alpha} \right) - \frac{1}{d - 1} \pi^{\mu \nu}\pi_{\alpha \beta}, \\&
	\Sigma^{\mu_i\nu_i}_{\alpha_i\beta_i}=\frac{p_{i\,\beta_i}}{p_i^2}\Big[2\delta^{(\nu_i}_{\alpha_i}p_i^{\mu_i)}-\frac{p_{i\alpha_i}}{(d-1)}\left(\delta^{\mu_i\nu_i}+(d-2)\frac{p_i^{\mu_i}p_i^{\nu_i}}{p_i^2}\right)\Big]+\frac{\pi^{\mu_i\nu_i}(p_i)}{(d-1)}\delta_{\alpha_i\beta_i}\equiv\mathcal{I}^{\mu_i\nu_i}_{\alpha_i}p_{i\,\beta_i} +\frac{\pi^{\mu_i\nu_i}(p_i)}{(d-1)}\delta_{\alpha_i\beta_i}.
\end{aligned}
\end{equation}
\normalsize
The link between the transverse and longitudinal and sectors is property of the $\langle AVV\rangle$ interaction at CFT level, while for this correlator this will not occur. We have summarised in appendix \ref{appendix:AVV} the approach in the case of the $\langle AVV\rangle$ in order to emphasize the difference between the two cases, which are remarkable. \\
With these information  at hand, the procedure to obtain the general structure of the correlator starts from the trace and conservation Ward identities
\begin{equation}
	\begin{aligned}
		&\nabla\cdot \langle J_c\rangle=0,\qquad\qquad && \nabla\cdot \langle J_5\rangle=a\,  \epsilon^{\mu \nu \rho \sigma}F_{\mu \nu}F_{\rho \sigma},\\&
		\nabla^\mu  \langle T_{\mu\nu}\rangle+F_{\mu\nu}\langle J^\mu\rangle+A_\nu\nabla_\mu \langle J^\mu\rangle=0, \qquad\qquad &&
		g_{\mu\nu}\langle T^{\mu\nu} \rangle=0.
	\end{aligned}
\end{equation}
We first assume a traceless energy-momentum tensor. Later we will also consider the case of an odd-parity trace anomaly.
Applying multiple functional derivatives to the Ward identities with respect to the field sources, after a Fourier transform, we find
\begin{equation}\label{eq:wijjt}
	\begin{aligned}
		&p_{i\mu_i} \langle J^{\mu_1}(p_1)J^{\mu_2}(p_2)T^{\mu_3 \nu_3}(p_3)\rangle=0,\quad \quad i=1,2,3\\
	\end{aligned}	
\end{equation}
		and
\begin{equation}\label{anntr}
		\begin{aligned}
		&\delta_{\mu_3 \nu_3} \langle J^{\mu_1}(p_1)J^{\mu_2}(p_2)T^{\mu_3 \nu_3}(p_3)\rangle=0.
	\end{aligned}
\end{equation}
Such equations are satisfied independently of the fact that $J$ are conserved vector or axial-vector currents. Notice that \eqref{anntr}, in the parity-even case would allow 2-point functions $JJ$ on the rhs, that here are, instead, absent. If the parity-odd operator of the $\langle JJT\rangle$ is $T$, then it is clear by symmetry that we cannot construct on the rhs a parity-odd $\langle JJ\rangle$ with two parity-even vector currents $J$. A similar result holds if both vector currents are parity-odd as well as $T$. If the two vector currents have mixed parity, say a $\langle J J\rangle_{odd}$, then one can easily check that this correlator vanishes. 

Due to the Ward identities \eqref{eq:wijjt}, the longitudinal part of the correlator vanishes. 
On the other hand, the transverse-traceless part can be formally expressed in terms of the following independent tensor structures and form factors
\begin{equation}\label{eq:decompJJT}
	\begin{aligned}
		\langle 
		J^{\mu_1}\left(p_1\right) J^{\mu_2}\left(p_2\right)&
		T^{\mu_3\nu_3}\left(p_3\right) 
		\rangle =
		\langle
		j^{\mu_1}\left(p_1\right) j^{\mu_2}	\left(p_2\right)
		t^{\mu_3\nu_3}\left(p_3\right) 
		\rangle =\\&
		\pi^{\mu_1}_{\alpha_1}
		\left(p_1\right) 
		\pi^{\mu_2}_{\alpha_2} \left(p_2\right)
		\Pi^{\mu_3\nu_3}_{\alpha_3\beta_3}\left(p_3\right)
		\Big[
		A_1\varepsilon^{p_1p_2\alpha_1\alpha_2}p_1^{\alpha_3}p_1^{\beta_3}+
		A_2\varepsilon^{p_1\alpha_1\alpha_2\alpha_3}p_1^{\beta_3}+\\&\hspace{6cm}
		A_3\varepsilon^{p_2\alpha_1\alpha_2\alpha_3}p_1^{\beta_3}
		+A_4\varepsilon^{p_1p_2\alpha_2\alpha_3}\delta^{\alpha_1\beta_3}
		\Big].
	\end{aligned}
\end{equation}
Note that we are not considering the following tensor structures in our decomposition
\begin{equation}
	p_2^{\alpha_1}p_1^{\beta_3}\varepsilon^{p_1p_2\alpha_2\alpha_3},\qquad
	p_3^{\alpha_2}p_1^{\beta_3}\varepsilon^{p_1p_2\alpha_1\alpha_3},\qquad
	\delta^{\beta_3\alpha_2}\varepsilon^{p_1p_2\alpha_1\alpha_3}.
\end{equation}
This is due to the Schouten identities \eqref{eq:JJTschoutid1} which relate these tensorial structures to those written in eq. \eqref{eq:decompJJT}. 
\subsection{Dilatation and Special Conformal Ward Identities}
In this section we start to analyse the conformal constraints on the form factors. The invariance of the correlator under dilatation is reflected in the equation
\begin{equation}
	\left(\sum_{i=1}^3 \Delta_i-2 d-\sum_{i=1}^2 p_i^\mu \frac{\partial}{\partial p_i^\mu}\right)\left\langle J^{\mu_1}\left(p_1\right) J^{\mu_2}\left(p_2\right) T^{\mu_3\nu_3}\left(p_3\right)\right\rangle=0.
\end{equation}
By using the chain rule
\begin{align}
	\frac{\partial}{\partial p_i^\mu}=\sum_{j=1}^3\frac{\partial p_j}{\partial p_i^\mu}\frac{\partial}{\partial p_j},
\end{align}
in term of the invariants $p_i=|\sqrt{p_i^2}|$ and by considering the decomposition \eqref{eq:decompJJT},
we can rewrite the dilatation equation as a constraint on the form factors
\begin{equation}
	\begin{aligned}
		\sum_{i=1}^{3} p_i \frac{\partial A_1}{\partial p_i }(p_1, p_2, p_3)-\left(\sum_{i=1}^{3} \Delta_i-2d-4 \right) A_1(p_1, p_2, p_3)=0\\
		\sum_{i=1}^{3} p_i \frac{\partial A_2}{\partial p_i }(p_1, p_2, p_3)-\left(\sum_{i=1}^{3} \Delta_i-2d-2\right) A_2(p_1, p_2, p_3)=0\\
		\sum_{i=1}^{3} p_i \frac{\partial A_3}{\partial p_i }(p_1, p_2, p_3)-\left(\sum_{i=1}^{3} \Delta_i-2d-2\right) A_3(p_1, p_2, p_3)=0\\
		\sum_{i=1}^{3} p_i \frac{\partial A_4}{\partial p_i }(p_1, p_2, p_3)-\left(\sum_{i=1}^{3} \Delta_i-2d-2 \right) A_4(p_1, p_2, p_3)=0.\\
	\end{aligned}
\end{equation}
The invariance of the correlator with respect to the special conformal transformations is instead encoded in the special conformal Ward identities 
\begin{align}\label{eq:scwijjtform}
	0=&\sum_{j=1}^2\left[-2\frac{\partial}{\partial p_{j\kappa}}-2p_j^\alpha\frac{\partial^2}{\partial p_j^\alpha\,\partial p_{j\kappa}}+p_j^\kappa\frac{\partial^2}{\partial p_j^\alpha\,\partial p_{j\alpha}}\right]\braket{J^{\mu_1}(p_1)J^{\mu_2}(p_2)T^{\mu_3\nu_3}(p_3)}\notag\\
	&+2\left(\delta^{\mu_1\kappa}\,\frac{\partial}{\partial p_1^{\alpha_1}}-\delta^\kappa_{\alpha_1}\,\frac{\partial}{\partial p_{1\mu_1}}\right)\braket{J^{\alpha_1}(p_1)J^{\mu_2}(p_2)T^{\mu_3\nu_3}(p_3)}\notag\\
	&+2\left(\delta^{\mu_2\kappa}\,\frac{\partial}{\partial p_2^{\alpha_2}}-\delta^\kappa_{\alpha_2}\,\frac{\partial}{\partial p_{2\mu_2}}\right)\braket{J^{\mu_1}(p_1)J^{\alpha_2}(p_2)T^{\mu_3\nu_3}(p_3)}\equiv \mathcal{K}^\kappa\braket{J^{\mu_1}(p_1)J^{\mu_2}(p_2)T^{\mu_3\nu_3}(p_3)}.
\end{align}
We can perform a transverse projection on all the indices in order to identify a set of independent partial differential equations  
\small
\begin{equation} \label{eq:decompJJTspec}
	\begin{aligned}
		&0=\pi_{\mu_1}^{\alpha_1}\left(p_1\right)
		\pi_{\mu_2}^{\alpha_2} \left(p_2\right) 
		\Pi_{\mu_3\nu_3}^{\alpha_3\beta_3} \left(p_3\right) 
		\mathcal{K}^k\left\langle J^{ \mu_1}\left(p_1\right) 
		J^{  \mu_2}\left(p_2\right) T^{\mu_3\nu_3}	\left(p_3\right)\right\rangle =\\ &
		\pi_{\mu_1}^{\alpha_1}\left(p_1\right)
		\pi_{\mu_2}^{\alpha_2} \left(p_2\right) 
		\Pi_{\mu_3\nu_3}^{\alpha_3\beta_3} \left(p_3\right) 
		\Biggl[
		\left(C_{11}\varepsilon^{p_1\alpha_1\alpha_2\alpha_3}p_1^{\beta_3}
		+C_{12}\varepsilon^{p_2\alpha_1\alpha_2\alpha_3}p_1^{\beta_3}
		+C_{13}\varepsilon^{p_1p_2\alpha_1\alpha_2}p_1^{\alpha_3}p_1^{\beta_3}
		+C_{14}\varepsilon^{p_1p_2\alpha_2\alpha_3}\delta^{\alpha_1\beta_3}\right)p_1^\kappa+\\ &
		\hspace{1.6cm}
		\left(C_{21}\varepsilon^{p_1\alpha_1\alpha_2\alpha_3}p_1^{\beta_3}
		+C_{22}\varepsilon^{p_2\alpha_1\alpha_2\alpha_3}p_1^{\beta_3}
		+C_{23}\varepsilon^{p_1p_2\alpha_1\alpha_2}p_1^{\alpha_3}p_1^{\beta_3}
		+C_{24}\varepsilon^{p_1p_2\alpha_2\alpha_3}\delta^{\alpha_1\beta_3}\right)p_2^\kappa+\\
		&\hspace{1.7cm}
		C_{31}\varepsilon^{\kappa \mu_1\mu_2\mu_3} p_1^{\nu_3}
		+C_{32}\varepsilon^{p_1\kappa \mu_2\mu_3}\delta^{\mu_1\nu_3}
		+C_{33}\varepsilon^{p_2\kappa \mu_1\mu_3}\delta^{\mu_2\nu_3}+
		C_{34}\varepsilon^{p_1p_2\kappa\mu_3}\delta^{\mu_1\mu_2}p_1^{\nu_3} +\\
		&\hspace{1.7cm}
		C_{41}\delta^{\mu_1 \kappa}\varepsilon^{\mu_2\mu_3p_1p_2} p_1^{\nu_3}
		+C_{51}\delta^{\mu_2 \kappa}\varepsilon^{\mu_1\mu_3p_1p_2} p_1^{\nu_3}
		+C_{61}	{\delta^{\mu_3 \kappa}\varepsilon^{p_1\mu_1\mu_2\nu_3}}
		+C_{62}{\delta^{\mu_3 \kappa}\varepsilon^{p_2\mu_1\mu_2\nu_3}}
		\Biggr],
	\end{aligned}
\end{equation}
\normalsize
where $C_{ij}$ are scalar functions of the form factors $A$ and their derivatives.
Such decomposition is obtained by writing all the possibile tensor structures and then identifying all the independent ones. The full procedure is shown in appendix \ref{appendix:Schouten}.
Due to the independence of the tensor structures of eq. \eqref{eq:decompJJTspec}, the special conformal constraints can then be written as
\begin{equation}
	C_{ij}=0\qquad i=1,\dots 6 \qquad j=1,\ldots 4. 
\end{equation}
\subsection{Solution of the CWIs}
In order to solve the CWIs, it is easier to first analyze the equations involving only the $A_1$ form factor
\begin{equation}\label{eq:a1cwis}
	\begin{aligned}
		K_{31}A_1=0, \qquad   
		K_{32}A_1=0, \qquad 
		\left(\frac{\partial }{\partial p_3}+4-\Delta_3\right)A_1=0.
	\end{aligned}
\end{equation}
The solution to the primary equations which are the first two is given by the following 3K integral
\begin{equation}
	A_1=b_1 J_{4\{0,0,0\}},
\end{equation}
which is not a divergent 3K integral in $d=4$ (see appendix \ref{appendix:3kint}), so in this case we can solve the last equation in \eqref{eq:a1cwis} directly without a regularization. We then set $\epsilon=0$ and $\Delta_3=4$. Performing the limit $p_1\rightarrow0$, we find the condition $b_1=0$. Therefore, we can write
\begin{equation}
	A_1=0.
\end{equation}
Inserting such solution back into the other conformal equations,
we can write the following primary special conformal Ward indentities
\begin{equation}
	\begin{aligned}
		&0=K_{31}A_2, \qquad    \qquad\qquad\qquad
		&&0=K_{32}A_2 -\left(\frac{2}{p_3}\frac{\partial}{\partial p_3}-\frac{2\Delta_3}{p_3^2}\right)\left(A_4+A_2-A_3\right), \\
		&0=K_{31}A_3  +\left(\frac{2}{p_3}\frac{\partial}{\partial p_3}-\frac{2\Delta_3}{p_3^2}\right)\left(A_4+A_2-A_3\right), \quad   
		&&0=K_{32}A_3,\\
		&0=K_{31}A_4, \qquad   
		&&0=K_{32}A_4.
	\end{aligned}
\end{equation}
These equations can be reduced to a set of homogenous equations by repeatedly applying the
operator $K_{ij}$ on them
\begin{equation}
	\begin{aligned}
		&0=K_{31}A_2, \qquad    \qquad\qquad\qquad
		&&0=K_{21}K_{32}A_2, \qquad, \\
		&0=K_{21}K_{31}A_3,  \qquad   
		&&0=K_{32}A_3,\\
		&0=K_{31}A_4, \qquad   
		&&0=K_{32}A_4, \\
	\end{aligned}
\end{equation}
which can be easly derived by looking at the non-homogenous equations and noticing that $K_{21}A_4=0$ and $K_{21}(A_2-A_3)=0$.
The solutions of the homogenous equations can then be written in terms of the following 3K integrals
\begin{equation}
	\begin{aligned}
		A_4= c_1 J_{2,\{0,0,0\}},\qquad\quad
		A_2=c_2 J_{3\{0,1,0\}}+c_3 J_{2,\{0,0,0\}},\qquad\quad
		A_3=c_4 J_{3\{1,0,0\}}+c_5 J_{2,\{0,0,0\}}.
	\end{aligned}
\end{equation} 
In $d=4+2 u \epsilon$, these integrals diverge like $1/\epsilon$ and therefore we should perform a regularization as we did for the other correlators. However here for simplicity we will avoid such procedure since such divergences don't particularly spoil our equations and we can arrive to the same conclusions with or without the regularization. Therefore, we will set
\begin{equation}
	d=4,\qquad\qquad\Delta_1=3,\qquad\qquad\Delta_2=3,\qquad\qquad\Delta_3=4.
\end{equation}
Inserting our solutions back into the non-homogeneous equations we find
\begin{equation}
	\begin{aligned}
		c_1=-4 c_2 - c_3 + c_5 ,\qquad\qquad
		c_4=-c_2. 
	\end{aligned}
\end{equation}
The secondary equations are given by $C_{ij}=0$ with $i\geq 3$. Their explicit expression is
\begin{equation}
	\begin{aligned}
		&0=p_2\frac{\partial A_2}{\partial p_2}+\frac{p_2^2+p_3^2}{p_3}\frac{\partial A_2}{\partial p_3}\\
		&0=p_2\frac{\partial A_3}{\partial p_2}-p_2\frac{\partial A_4}{\partial p_2}+\frac{p_2^2+p_3^2}{p_3}\frac{\partial A_3}{\partial p_3}\\
		&0= -\frac{8}{p_3^2}A_2+\left(\frac{4}{p_1^2}+\frac{8}{p_3^2}\right)A_3-\left(\frac{4}{p_1^2}+\frac{8}{p_3^2}\right)A_4-\frac{2p_1}{p_3^2}\frac{\partial A_2}{\partial p_1}\\
		&\qquad\quad
		+\left(-\frac{2}{p_1}+\frac{2 p_1}{p_3^2}\right)\frac{\partial A_3}{\partial p_1}
		+\left(\frac{2}{p_1}-\frac{2 p_1}{p_3^2}\right)\frac{\partial A_4}{\partial p_1}
		-\frac{2p_2}{p_3^2}\frac{\partial A_2}{\partial p_2}
		+\frac{2p_2}{p_3^2}\frac{\partial A_3}{\partial p_2}
		-\frac{2p_2}{p_3^2}\frac{\partial A_4}{\partial p_2}
		,\\
		&0=\left(\frac{4}{p_2^2}+\frac{8}{p_3^2}\right)A_2-\frac{8}{p_3^2}A_3+\frac{8}{p_3^2}A_4
		-\frac{2}{p_2}\frac{\partial A_2}{\partial p_2}
		-\frac{2}{p_3}\frac{\partial A_2}{\partial p_3}
		+\frac{2}{p_3}\frac{\partial A_3}{\partial p_3}
		-\frac{2}{p_3}\frac{\partial A_4}{\partial p_3}\\
		&0=A_2+A_3-A_4-p_2\frac{\partial A_2}{\partial p_2}
		+p_2\frac{\partial A_3}{\partial p_2} +p_2\frac{\partial A_4}{\partial p_2} 
		-p_3\frac{\partial A_2}{\partial p_3} +p_3\frac{\partial A_4}{\partial p_3} 
		\\
		&0=A_3-A_4+p_2\frac{\partial A_4}{\partial p_2}+p_3\frac{\partial A_4}{\partial p_3}\\
		&0=A_2+{p_2}\frac{\partial A_4}{\partial p_2}.
	\end{aligned}
\end{equation}
Working in the limit $p_3\rightarrow0$, the first and second equation lead to the following constraints
\begin{equation}
	c_3=-2c_2, \qquad c_5=4c_2.
\end{equation}
Inserting such conditions into the last equation and using the properties \eqref{eq:3kprop} of the 3K integrals, we arrive to $c_2=0$.
In the end we find that the correlator is identically null
\begin{equation}
	\begin{aligned}
		\langle
		J^{\mu_1}\left(p_1\right) J^{\mu_2}&	\left(p_2\right)
		T^{\mu_3\nu_3}\left(p_3\right) 
		\rangle =0.
	\end{aligned}
\end{equation}
We conclude that in the absence of a parity-odd trace anomaly, the parity-odd $\langle JJT\rangle $ vanishes. This result is 
slightly modified when we include a parity-odd trace anomaly, with the result expressed uniquely in terms of an anomaly pole, as anticipated in \eqref{pp1}. We are going to show this in the next section. 

\section{The $\langle JJT \rangle $ correlator in CFT with an anomalous trace}
In the previous section we have assumed that there is no parity-odd trace anomaly. In this section we  relax such assumption by considering a term of the following form
\begin{equation}
	g_{\mu\nu}\langle T^{\mu\nu} \rangle=f \, \varepsilon^{\mu\nu\rho\sigma} F_{\mu\nu}F_{\rho\sigma},
\end{equation} 
where $f$ is an arbitrary constant.
In this case, the local part of the $\langle JJT\rangle$ correlator does not vanish anymore. Indeed, applying multiple functional derivatives to such equations with respect to the field sources, after a Fourier transform, we find
\begin{equation}\label{eq:jjttracciann}
	g_{\mu\nu}\langle J^{\mu_1}(p_1)J^{\mu_2}(p_2)T^{\mu\nu} (p_3)\rangle=8\, f\, \varepsilon^{p_1p_2\mu_1\mu_2}.
\end{equation}
Therefore, there will be a non-vanishing term of the form $\left\langle  j^{\mu_1} j^{\mu_2}t_{l o c}^{\mu_3 \nu_3}\right\rangle$ contributing to the correlator. Note that we are still assuming the conservation of the currents and the energy-momentum tensor \cite{Bonora:2021mir}
\begin{equation}
	p_{i\mu_i} \langle J^{\mu_1}(p_1)J^{\mu_2}(p_2)T^{\mu_3 \nu_3}(p_3)\rangle=0,\qquad \qquad i=1,2,3.
\end{equation}
Using eq. \eqref{eq:jjttracciann}, we can then write the correlator in terms of its longitudinal and transverse-traceless parts 
\begin{equation}
	\begin{aligned}
		\langle
		J^{\mu_1}\left(p_1\right) J^{\mu_2}&	\left(p_2\right)
		T^{\mu_3\nu_3}\left(p_3\right) 
		\rangle =\left\langle  j^{\mu_1} (p_1)j^{\mu_2}(p_2)t_{l o c}^{\mu_3 \nu_3}(p_3)\right\rangle+
		\left\langle  j^{\mu_1} (p_1)j^{\mu_2}(p_2)t^{\mu_3 \nu_3}(p_3)\right\rangle,
	\end{aligned}
\end{equation}
with
\begin{equation}
	\left\langle  j^{\mu_1} (p_1)j^{\mu_2}(p_2)t_{l o c}^{\mu_3 \nu_3}(p_3)\right\rangle=
	\frac{8}{3}\,f\, \pi^{\mu_3\nu_3}(p_3) \varepsilon^{p_1p_2\mu_1\mu_2}.
\end{equation}
In order to fix the transverse-traceless part, we need to analyze the conformal constraints in the presence of an anomaly. As we have seen in section 3, the conformal equations on the $\langle JJT \rangle $ are not violated by the trace anomaly $F\tilde{F}$. 
Therefore we can apply the same procedure discussed in the previous section for the traceless case, but with the addition of a term of the form $\left\langle  j^{\mu_1} j^{\mu_2}t_{l o c}^{\mu_3 \nu_3}\right\rangle$ contributing to the correlator.
However such term does not affect the conformal constraints written before for the transverse-traceles part since
\begin{equation}
	\pi_{\mu_1}^{\rho_1}\left(p_1\right) \pi_{\mu_2}^{\rho_2}\left(p_2\right) 
	\Pi_{\mu_3\nu_3}^{\rho_3\sigma_3} \left(p_3\right)
	K^\kappa
	\left\langle  j^{\mu_1} j^{\mu_2}t_{l o c}^{\mu_3 \nu_3}\right\rangle
	=0.
\end{equation}
We can prove such equation by using the conservation of the energy-momentum tensor together with the properties of the projectors.
Therefore, our analysis of the transverse-traceless part in the previous section still applies.
The transverse-traceless part is still zero and the correlator is purely longitudinal.\\ 
We now recall the definition of the local term
\begin{equation}
	t_{l o c}^{\mu \nu}(p)\equiv \left(I_{\alpha \beta}^{\mu \nu}+\frac{1}{d-1} \pi^{\mu \nu} \delta_{\alpha \beta}\right) T^{\alpha \beta}(p)=\frac{1}{d-1} \pi^{\mu \nu}  T^{\alpha}_\alpha(p),
\end{equation}
where in the second identity relation we have used the conservation of the energy-momentum tensor.
We can then write the correlator purely in terms of its local part
\begin{equation}
	\begin{aligned}
		\langle
		J^{\mu_1}\left(p_1\right) J^{\mu_2}&	\left(p_2\right)
		T^{\mu_3\nu_3}\left(p_3\right) 
		\rangle =\left\langle  J^{\mu_1} (p_1)J^{\mu_2}(p_2)t_{l o c}^{\mu_3 \nu_3}(p_3)\right\rangle=
		\frac{8}{3}\,f\, \pi^{\mu_3\nu_3}(p_3) \varepsilon^{p_1p_2\mu_1\mu_2}.
	\end{aligned}
\end{equation}
We have found that the parity-odd structure of the correlator is forced to appear only in its trace sector with the transverse traceless sector being identically zero. \\
As further proof, one can check by a direct computation that this solution satisfies the non anomalous CWIs \eqref{DJJT} and \eqref{SJJT}.

\section{The $\langle TTO\rangle$ correlator in CFT}
In this section we study the conformal constraints on the $\braket{TTO}$ correlator where each $T$ can be the standard energy-momentum tensor or $T_5$ of odd parity.
Our result is valid for every odd parity $\braket{TTO}$ correlator constructed with standard energy-momentum tensors and/or $T_5$ and a scalar/pseudoscalar operator $O$. As we will see, also in this case the solution of the CWIs of the correlator can be written in terms of 3K integrals that needs a regularization. Therefore, we work in the general scheme $\{ u,v\}$ where $d=4+2 u \epsilon $, with the conformal dimensions $\Delta_i$ of the operators is shifted by $(u+v_i) \, \epsilon $ with arbitrary $u$ and $v_i$.\\
We can procede in a manner similar to the previous correlators by decomposing the $\langle TTO\rangle$ into its longitudinal and transverse-traceless parts. 
In this case, the conservation and trace Ward identities for the energy-momentum tensor lead to the condition\footnote{
	Note that such equations remain valid even if we allow an odd parity term $\varepsilon^{\mu \nu \rho \sigma}R^{\alpha\beta}_{\hspace{0.3cm} \mu \nu} R_{\alpha\beta \rho \sigma}$ in the trace anomaly. Indeed if we consider the case $O=T^\mu_\mu$ and we trace again over one of the other energy-momentum tensors, we get a vanishing result. We will show this more in detail when examining the $\langle TTT\rangle$ correlator.}
\begin{equation} \label{eq:conswittoa}
	p_{i\mu_i} \left\langle T^{ \mu_1\nu_1}\left(p_1\right) 
	\right.\left.
	T^{ \mu_2\nu_2}\left(p_2\right) O	\left(p_3\right)\right\rangle =0, \qquad\qquad 
	g_{\mu_i\nu_i} \left\langle T^{ \mu_1\nu_1}\left(p_1\right) 
	\right.\left.
	T^{ \mu_2\nu_2}\left(p_2\right) O	\left(p_3\right)\right\rangle =0,
\end{equation}
for $i=\{1,2\}$. 
Due to the equations \eqref{eq:conswittoa}, the longitudinal part of the correlator vanishes.
On the other hand, the transverse-traceless part can be formally expressed in terms of the following form factors
\begin{equation}
	\begin{aligned}
		\left\langle T^{ \mu_1\nu_1}\left(p_1\right) 
		\right.&\left.
		T^{ \mu_2\nu_2}\left(p_2\right) O	\left(p_3\right)\right\rangle =\left\langle t^{ \mu_1\nu_1}\left(p_1\right) 
		\right.\left.
		t^{ \mu_2\nu_2}\left(p_2\right) O	\left(p_3\right)\right\rangle =
		\\
		&\Pi^{\mu_1\nu_1}_{\alpha_1\beta_1}\left(p_1\right)
		\Pi^{\mu_2\nu_2}_{\alpha_2\beta_2} \left(p_2\right) \Bigl[  A_1(p_1,p_2,p_3)\varepsilon^{\alpha_1\alpha_2 p_1 p_2}p^{\beta_1}_2p_3^{\beta_2}
		+A_2(p_1,p_2,p_3)\varepsilon^{\alpha_1\alpha_2 p_1 p_2}\delta^{\beta_1\beta_2} 
		\Bigr].
	\end{aligned}
\end{equation}
\subsection{Dilatation and Special Conformal Ward Identities} \label{secEqTTO}
We can now analyse the conformal constraints on the form factors $A_1$ and $A_2$. We proceed in a manner similar to the previous correlators. The invariance of the $\langle TTO \rangle$ under dilatation is reflected in the following constraints on the form factors
\begin{equation}
	\begin{aligned}
		\sum_{i=1}^{3} p_i \frac{\partial A_1}{\partial p_i }-\left(\sum_{i=1}^3\Delta_i-2d- 4\right) A_1=0,\\
		\sum_{i=1}^{3} p_i \frac{\partial A_2}{\partial p_i }-\left(\sum_{i=1}^3\Delta_i-2d- 2\right) A_2=0.
	\end{aligned}
\end{equation}
The invariance of the correlator with respect to the special conformal transformations is instead encoded in the special conformal Ward identities 
\begin{equation}
	\begin{aligned}
		0=  \mathcal{K}^{{\kappa}}&\left\langle T^{\mu_1 \nu_1}\left(p_1\right) T^{\mu_2 \nu_2}\left(p_2\right) O\left(p_3\right)\right\rangle =\\
		\sum_{j=1}^2\bigg(  & 2\left(\Delta_j-d\right) \frac{\partial}{\partial p_{j \kappa}}-2 p_j^\alpha \frac{\partial}{\partial p_j^\alpha} \frac{\partial}{\partial p_{j \kappa}}+\left(p_j\right)^\kappa \frac{\partial}{\partial p_j^\alpha} \frac{\partial}{\partial p_{j \alpha}} \bigg) \left\langle T^{\mu_1 \nu_1}\left(p_1\right) T^{\mu_2 \nu_2}\left(p_2\right) O\left(p_3\right)\right\rangle \\
		& +4\left(\delta^{\kappa\left(\mu_1\right.} \frac{\partial}{\partial p_1^{\alpha_1}}-\delta_{\alpha_1}^\kappa \delta_\lambda^{\left(\mu_1\right.} \frac{\partial}{\partial p_{1 \lambda}}\right)\left\langle T^{\left.\nu _1\right) \alpha_1}\left(p_1\right) T^{\mu_2 \nu_2}\left(p_2\right) O\left(p_3\right)\right\rangle \\
		& +4\left(\delta^{\kappa\left(\mu_2\right.} \frac{\partial}{\partial p_2^{\alpha_2}}-\delta_{\alpha_2}^\kappa \delta_\lambda^{\left(\mu_2\right.} \frac{\partial}{\partial p_{2 \lambda}}\right)\left\langle T^{\left.\nu_2\right) \alpha_2}\left(p_2\right) T^{\mu_1 \nu_1}\left(p_1\right) O\left(p_3\right)\right\rangle .
	\end{aligned}
\end{equation}
We can perform a transverse projection on all the indices in order to identify a set of partial differential equations, using the following minimal decomposition  
\begin{equation}
\label{all}
	\begin{aligned}
		0=&\Pi_{\mu_1\nu_1}^{\rho_1\sigma_1}\left(p_1\right)
		\Pi_{\mu_2\nu_2}^{\rho_2\sigma_2} \left(p_2\right) 
		\mathcal{K}^k\left\langle T^{\mu_1 \nu_1}\left(p_1\right) 
		T^{\mu_2\nu_2}\left(p_2\right) O	\left(p_3\right)\right\rangle =
		\Pi_{\mu_1\nu_1}^{\rho_1\sigma_1}\left(p_1\right)
		\Pi_{\mu_2\nu_2}^{\rho_2\sigma_2} \left(p_2\right) \Bigl[\\
		&\hspace{1.5cm}
		C_{11}\varepsilon^{\mu_1\mu_2 p_1 p_2}p^{\nu_1}_2 p^{\nu_2}_3p_1^\kappa
		+	C_{12}\varepsilon^{\mu_1\mu_2 p_1 p_2}\delta^{\nu_1\nu_2}_3p_1^\kappa
		+C_{21}\varepsilon^{\mu_1\mu_2 p_1 p_2}p^{\nu_1}_2 p^{\nu_2}_3 p_2^\kappa
		+C_{22}\varepsilon^{\mu_1\mu_2 p_1 p_2}\delta^{\nu_1 \nu_2}_3 p_2^\kappa
		\\&\hspace{1.4cm}
		+C_{31}\varepsilon^{p_1\kappa\mu_1\mu_2}\delta^{\nu_1\nu_2}
		+C_{32}\varepsilon^{p_2\kappa\mu_1\mu_2}\delta^{\nu_1\nu_2}
		+C_{41} \varepsilon^{p_1 p_2 \mu_1 \mu_2}p_3^{\nu_2} \delta^{\kappa \nu_1}
		+C_{51} \varepsilon^{p_1 p_2 \mu_1 \mu_2} p_2^{\nu_1}\delta^{\kappa \nu_2}
		\Bigr].
	\end{aligned}
\end{equation}
All the tensor structures in this formula are independent. Indeed, we have not considered the following tensors
\begin{equation}
	\begin{aligned}
		&\varepsilon^{p_1 \kappa \mu_1 \mu_2 }p^{\nu_1}_2p_3^{\nu_2},
		\qquad \qquad
		&&\varepsilon^{p_2 \kappa \mu_1 \mu_2 }p^{\nu_1}_2p_3^{\nu_2},
		\qquad \qquad
		&&\varepsilon^{p_1 p_2 \kappa\mu_2 }p_2^{\mu_1}p^{\nu_1}_2p_3^{\nu_2},\\
		&\varepsilon^{p_1 p_2 \kappa\mu_1 }p_3^{\mu_2}p^{\nu_1}_2p_3^{\nu_2},
		&&\varepsilon^{p_1 p_2\kappa \mu_1 } p_3^{\nu_2}\delta^{\mu_2\nu_1},
		&&\varepsilon^{p_1p_2\kappa\mu_2}p_2^{\nu_1}\delta^{\mu_1\nu_2},
	\end{aligned}
\end{equation}
which can be rewritten in terms of the ones present in our decomposition, using the Schouten identities
\begin{equation}
	\begin{aligned}
		\epsilon^{[p_1 p_2 \mu_1 \mu_2} p_1^{\kappa]}=0\\
		\epsilon^{[p_1 p_2 \mu_1 \mu_2} p_2^{\kappa]}=0\\
		\epsilon^{[p_1 p_2 \mu_1 \mu_2} \delta^{\kappa]\nu_1}=0\\
		\epsilon^{[p_1 p_2 \mu_1 \mu_2} \delta^{\kappa]\nu_2}=0,
	\end{aligned}
\end{equation}
according to which we have
\small
\begin{equation}
	\begin{aligned} 
		\Pi_{\mu_1\nu_1}^{\rho_1\sigma_1 }(p_1)\Pi_{\mu_2\nu_2}^{\rho_2\sigma_2} (p_2)\bigg(
		\epsilon^{p_1 p_2 {\kappa}{\mu_1}} p_3^{\mu_2}\bigg)&=
		\Pi_{\mu_1\nu_1}^{\rho_1\sigma_1 }(p_1)\Pi_{\mu_2\nu_2}^{\rho_2\sigma_2} (p_2)\bigg(
		\frac{1}{2} \epsilon^{p_1{\kappa}{\mu_1}{\mu_2}} (p_1^2+p_2^2-p_3^2)+\epsilon^{p_1p_2{\mu_1}{\mu_2}} p_1^{\kappa}+\epsilon^{p_2{\kappa}{\mu_1}{\mu_2}} p_1^2
		\bigg)
		\\
		\Pi_{\mu_1\nu_1}^{\rho_1\sigma_1 }(p_1)\Pi_{\mu_2\nu_2}^{\rho_2\sigma_2} (p_2)\bigg(
		\epsilon^{p_1p_2{\kappa}{\mu_2}} p_2^{\mu_1}\bigg)&=
		\Pi_{\mu_1\nu_1}^{\rho_1\sigma_1 }(p_1)\Pi_{\mu_2\nu_2}^{\rho_2\sigma_2} (p_2)\bigg(
		-\frac{1}{2} \epsilon^{p_2{\kappa}{\mu_1}{\mu_2}} (p_1^2+p_2^2-p_3^2)
		+\epsilon^{p_1p_2{\mu_1}{\mu_2}} p_2^{\kappa}
		-\epsilon^{p_1{\kappa}{\mu_1}{\mu_2}} p_2^2\bigg)\\
		\Pi_{\mu_1\nu_1}^{\rho_1\sigma_1 }(p_1)\Pi_{\mu_2\nu_2}^{\rho_2\sigma_2} (p_2)\bigg(
		\varepsilon^{\mu_1\mu_2\kappa p_1}p_2^{ \nu_1}\bigg)&=
		\Pi_{\mu_1\nu_1}^{\rho_1\sigma_1 }(p_1)\Pi_{\mu_2\nu_2}^{\rho_2\sigma_2} (p_2)\bigg(
		-\varepsilon^{\kappa p_1 p_2 \mu_1}\delta^{\mu_2\nu_1}
		- \varepsilon^{p_1p_2\mu_1\mu_2}\delta^{\kappa \nu_1}\bigg)\\
		\Pi_{\mu_1\nu_1}^{\rho_1\sigma_1}(p_1)\Pi_{\mu_2\nu_2}^{\rho_2\sigma_2} (p_2)\bigg(
		\varepsilon^{p_2 \mu_1\mu_2\kappa }p_1^{ \nu_2}\bigg)&=
		\Pi_{\mu_1\nu_1}^{\rho_1\sigma_1 }(p_1)\Pi_{\mu_2\nu_2}^{\rho_2\sigma_2} (p_2)\bigg(
		-\varepsilon^{\mu_2 \kappa p_1 p_2}\delta^{\mu_1\nu_2}
		- \varepsilon^{p_1p_2\mu_1\mu_2}\delta^{\kappa \nu_2}\bigg).\\
	\end{aligned}
\end{equation}
\normalsize
We remark that we can always change $(\mu_1\leftrightarrow \nu_1)$ and/or $(\mu_2\leftrightarrow \nu_2)$ in order to obtain new Schouten identities.
Due to the independence of tensor structures in our decomposition, the special conformal equations are written, from \eqref{all}, as 
\beq
C_{ij}=0\qquad i=1,\ldots 4 \qquad j=1,2.
\eeq
The equations with $i=\{1,2\}$ are second order differential equations called primary equations. Their explicit form is
\begin{equation}\label{eq:primaryTTOn}
	\begin{aligned}
		&K_{31}A_1=0,\qquad 
		&&K_{32}A_1=0,\\
		&K_{31}A_2-2p_2 \frac{\partial A_1}{\partial p_2 }+4 A_1=0, \qquad
		&&K_{32}A_2-2p_1 \frac{\partial A_1}{\partial p_1 }+4 A_1=0.
	\end{aligned}
\end{equation}
The other equations are of the first order and are called secondary ones. Their explicit form is 
\begin{equation} \label{eq:seceqtto}
	\begin{aligned}
		&0=(p_1^2-p_2^2-p_3^2)A_1+2A_2+2p_1p_2^2\frac{\partial A_1}{\partial p_1}-p_2(p_1^2+p_2^2-p_3^2)\frac{\partial A_1}{\partial p_2}-2p_2\frac{\partial A_2}{\partial p_2}, \\
		&0=(p_1^2-p_2^2+p_3^2)A_1-2A_2-2p_1^2p_2\frac{\partial A_1}{\partial p_2}+p_1(p_1^2+p_2^2-p_3^2)\frac{\partial A_1}{\partial p_1}+2p_1\frac{\partial A_2}{\partial p_1}, \\
		&0=2\left(  \frac{p_1^2+2p_2^2-2p_3^2}{p_1^2}  \right) A_1+\frac{8}{p_1^2}A_2 -\frac{p_1^2+p_2^2-p_3^2}{p_1}
		\frac{\partial A_1}{\partial p_1}-\frac{2}{p_1}\frac{\partial A_2}{\partial p_1}-4p_2\frac{\partial A_1}{\partial p_2},\\
		&0=-2\left( \frac{2p_1^2+p_2^2-2p_3^2}{p_2^2}\right)A_1-\frac{8}{p_2^2}A_2+
		4p_1\frac{\partial A_1}{\partial p_1}+\frac{p_1^2+p_2^2-p_3^2}{p_2}\frac{\partial A_1}{\partial p_2}+\frac{2}{p_2}\frac{\partial A_2}{\partial p_2}.
	\end{aligned}
\end{equation}

\subsection{Solutions of the CWIs}\label{secSolTTO}
In order to solve the primary eqs$.$ \eqref{eq:primaryTTOn}, we rewrite them as a set of homogeneous equations by repeatedly applying the
operator $K_{ij}$ on them
\begin{equation}
	\begin{aligned}
		&K_{31}A_1=0,\qquad 
		&&K_{32}A_1=0,\\
		&K_{31}K_{31}A_2=0, \qquad
		&&K_{32}K_{32}A_2=0.
	\end{aligned}
\end{equation}
The most general solution of these equations can be written in terms of the following combinations of 3K integrals
\begin{equation}
	\begin{aligned} \label{eq:solomogtto}
		&A_1=c_1 J_{4\{0,0,0\}},\\
		&A_2=c_2 J_{3\{1,0,0\}}+c_3 J_{3\{0,1,0\}}+c_4 J_{3\{0,0,1\}}+c_5 J_{4\{1,1,0\}}+c_6J_{2\{0,0,0\}}.
	\end{aligned}
\end{equation} 
We then insert these solutions back into the non-homogeneous primary eqs$.$ \eqref{eq:primaryTTOn} and the secondary eqs$.$ \eqref{eq:seceqtto} in order to fix the constants $c_i$. We can solve such constraints for different values of the conformal dimensions $\Delta_3$.
As seen in the computations of the previous correlators, the procedure may involve a regularization,  the use of the properties of 3K integrals and their limits $p_i\rightarrow0$ described in the Appendix \ref{appendix:3kint}. For odd values of $\Delta_3$, no regularization is needed, and we found only vanishing solutions. We also considered different examples with even values of $\Delta_3$. In particular, we found
	\begin{align}
			&
		\left\langle T^{ \mu_1\nu_1}\left(p_1\right) 
		T^{ \mu_2\nu_2}\left(p_2\right) O_{\left(\Delta_3=-2 \right)}	\left(p_3\right)\right\rangle =
		\Pi^{\mu_1\nu_1}_{\alpha_1\beta_1}\left(p_1\right)
		\Pi^{\mu_2\nu_2}_{\alpha_2\beta_2} \left(p_2\right)  \,  \times\nonumber \\ &\qquad \frac{c_1}{p_3^8}\Biggl[  
		\left(3\left(p_1^2-p_2^2\right)^2-2\left(p_1^2+p_2^2\right)p_3^2-p_3^4\right)
		\varepsilon^{\alpha_1\alpha_2 p_1 p_2}\delta^{\beta_1\beta_2} 
		-2\left(3p_1^2+3p_2^2+p_3^2\right)\varepsilon^{\alpha_1\alpha_2 p_1 p_2}p^{\beta_1}_2p_3^{\beta_2}
		\Biggr],\nonumber \\[8pt]
		&
		\left\langle T^{ \mu_1\nu_1}\left(p_1\right) 
		T^{ \mu_2\nu_2}\left(p_2\right) O_{\left(\Delta_3=0 \right)}	\left(p_3\right)\right\rangle =
		\nonumber\\ &\qquad\qquad
		\Pi^{\mu_1\nu_1}_{\alpha_1\beta_1}\left(p_1\right)
		\Pi^{\mu_2\nu_2}_{\alpha_2\beta_2} \left(p_2\right) \, \frac{c_1}{p_3^4}\Biggl[  
		-\frac{p_1^2+p_2^2-p_3^2}{2}
		\varepsilon^{\alpha_1\alpha_2 p_1 p_2}\delta^{\beta_1\beta_2} 
		+\varepsilon^{\alpha_1\alpha_2 p_1 p_2}p^{\beta_1}_2p_3^{\beta_2}
		\Biggr],\nonumber\\[8pt]
		&
		\left\langle T^{ \mu_1\nu_1}\left(p_1\right) 
		T^{ \mu_2\nu_2}\left(p_2\right) O_{\left(\Delta_3=2 \right)}	\left(p_3\right)\right\rangle =0
		,\nonumber\\[8pt] 
		&
		\left\langle T^{ \mu_1\nu_1}\left(p_1\right) 
		T^{ \mu_2\nu_2}\left(p_2\right) O_{\left(\Delta_3=4 \right)}	\left(p_3\right)\right\rangle =\nonumber
		\\ &\qquad\qquad
		\Pi^{\mu_1\nu_1}_{\alpha_1\beta_1}\left(p_1\right)
		\Pi^{\mu_2\nu_2}_{\alpha_2\beta_2} \left(p_2\right) \, c_1\Biggl[  
		-\frac{p_1^2+p_2^2-p_3^2}{2}
		\varepsilon^{\alpha_1\alpha_2 p_1 p_2}\delta^{\beta_1\beta_2} 
		+\varepsilon^{\alpha_1\alpha_2 p_1 p_2}p^{\beta_1}_2p_3^{\beta_2}
		\Biggr],\nonumber\\[8pt]
		&
		\left\langle T^{ \mu_1\nu_1}\left(p_1\right) 
		T^{ \mu_2\nu_2}\left(p_2\right) O_{\left(\Delta_3=6 \right)}	\left(p_3\right)\right\rangle =
		\Pi^{\mu_1\nu_1}_{\alpha_1\beta_1}\left(p_1\right)
		\Pi^{\mu_2\nu_2}_{\alpha_2\beta_2} \left(p_2\right)  \,  \times \nonumber\\ &\qquad c_1\Biggl[  
		\left(3\left(p_1^2-p_2^2\right)^2-2\left(p_1^2+p_2^2\right)p_3^2-p_3^4\right)
		\varepsilon^{\alpha_1\alpha_2 p_1 p_2}\delta^{\beta_1\beta_2} 
			-2\left(3p_1^2+3p_2^2+p_3^2\right)\varepsilon^{\alpha_1\alpha_2 p_1 p_2}p^{\beta_1}_2p_3^{\beta_2}
		\Biggr],\nonumber\\[8pt]
		&
		\left\langle T^{ \mu_1\nu_1}\left(p_1\right) 
		T^{ \mu_2\nu_2}\left(p_2\right) O_{\left(\Delta_3=8 \right)}	\left(p_3\right)\right\rangle =0.
	\end{align}
\normalsize
Note that we have also considered examples with $\Delta_3 \leq 0$. Although these cases violate unitarity and are non-physical, they can be relevant in particular contexts such as holography. It is important to note that the conformal equations do not require these non-physical solutions to vanish.\\
We now focus on the solution with $\Delta_3=4$, which is satisfied for example if $O=\nabla\cdot J_5$ or $O=T^\mu_\mu$. After contracting the indices of the transverse-traceless projectors with the argument in the square brackets, we have
\begin{equation}
	\begin{aligned}
		&
		\left\langle T^{ \mu_1\nu_1}\left(p_1\right) 
		T^{ \mu_2\nu_2}\left(p_2\right) O_{\left(\Delta_3=4 \right)}	\left(p_3\right)\right\rangle =
		\\ 
		&\qquad\qquad
		\frac{c_1}{4}\Biggl[ \varepsilon^{\nu_1\nu_2 p_1 p_2} \bigg((p_1\cdot p_2)\delta^{\mu_1\mu_2}-p_{1}^{\mu_2}p_{2}^{\mu_1}\bigg)+\left(\mu_1\leftrightarrow\nu_1\right)+\left(\mu_2\leftrightarrow\nu_2\right)
		+\left(\begin{aligned}\mu_1\leftrightarrow\nu_1\\ \mu_2\leftrightarrow\nu_2\end{aligned}\right)\Biggr].
	\end{aligned}
\end{equation}
This solution
can be rewritten as functional derivatives of $R\tilde{R}$ 
\begin{equation}
	\begin{aligned}
		&\delta^4(p_1+p_2+p_3)\left\langle T^{ \mu_1\nu_1}\left(p_1\right) 
		T^{ \mu_2\nu_2}\left(p_2\right) O_{(\Delta_3= 4)}	\left(p_3\right)\right\rangle =  \\&
		\qquad\qquad \int  dx_1\, dx_2 \,dx_3\, e^{-i(p_1x_1+p_2x_2+p_3x_3)}\, \, \frac{\delta^2   \Big[f \, \varepsilon^{\mu \nu\rho\sigma}R^{\alpha\beta}_{\hspace{0.3cm} \mu \nu}(x_3)R_{\alpha\beta \rho \sigma}(x_3)\Big] }{\delta g_{\mu_1\nu_1}(x_1)\,\, \, \delta g_{\mu_2\nu_2} (x_2)}
	\end{aligned}
\end{equation}
in agreement with the chiral anomaly formula \eqref{eq:anomaliachirale} in the case $O_{(\Delta_3= 4)}	=\nabla \cdot J_5$ and potentially a parity-odd trace anomaly $R \tilde{R}$ for the case $O_{(\Delta_3= 4)}	=T^\mu_\mu$.

\section{The $\langle TTT\rangle$ correlator in CFT}
This correlator has been studied in several previous works in coordinate space in a conformal field theory and it was found by Stanev, for instance, that its parity-odd sector vanishes \cite{Stanev:2012nq}. The analysis in coordinate space avoids the contact terms where all the external points of the correlator coalesce, which are the source of the conformal anomaly.\footnote{The inclusion of such contact terms, in the parity even case, has been discussed instead long ago in \cite{Osborn:1993cr}. } \\
However, in these investigations it was always assumed that odd-parity trace terms were absent
\begin{equation} \label{eq:notracettt}
	g_{\mu_i\nu_i}\langle T^{\mu_1\nu_1}T^{\mu_2\nu_2}T^{\mu_3\nu_3}\rangle_{odd}=0.
\end{equation}
Moreover, this correlator vanishes in all the perturbative analysis presented so far, in different schemes, as for instance in \cite{Bonora:2022izj}. Indeed, the claim in \cite{Bonora:2022izj} is that the gravitational trace anomaly comes entirely from the second term of  \eqref{eq:defanomduff}.
Since our goal is to investigate only the critical cases raised by the perturbative analysis, we will not investigate completely all the sectors of the $\langle TTT\rangle_{odd}$ when parity-odd trace anomalies are present.

We are going to show that while all the perturbative analysis indicate that the  $\langle TTT \rangle_{odd}$  is identically zero and that there is no parity-odd trace anomaly from the perturbative standpoint, if we allow a parity-odd trace anomaly, where the trace is performed after the quantum average of $T$, then the same correlator is not predicted to be zero by the CWIs in its longitudinal and trace sectors. The result clearly differs from the ordinary perturbative analysis presented before. Notice that explicit structure of the transverse/traceless sector, which is unknown, is irrelevant for this goal, since the three sectors are 
orthogonal.  \\
If we admit the Pontryagin density $f \,\varepsilon^{\mu\nu \rho\sigma}R_{\alpha\beta\mu\nu}R^{\alpha\beta}_{\,\,\,\,\,\rho\sigma}$ as an anomalous trace term for the correlator, we can write
\begin{equation}
	\delta_{\mu_3\nu_3} \left\langle T^{\mu_1 \nu_1} T^{\mu_2\nu_2} T^{ \mu_3\nu_3}\right\rangle= -16  \, f  \, \Bigg\{ \bigg[\varepsilon^{\nu_1 \nu_2 p_1 p_2}\bigg( (p_1 \cdot p_2) \delta^{\mu_1 \mu_2}- {p_1^{\mu_2} p_2^{\mu_1}}\bigg) +\left( \mu_1 \leftrightarrow \nu_1 \) \bigg] +\left( \mu_2 \leftrightarrow \nu_2 \right) \Bigg\}.
\end{equation}
We can derive analogous expression with the trace over $T^{\mu_1 \nu_1}$ or $T^{\mu_2 \nu_2}$ by exchanging the momenta and indices $\{p_{1}\leftrightarrow p_3,\, \mu_{1}\leftrightarrow \mu_3\}$ or $\{p_{2}\leftrightarrow p_3,\, \mu_{2}\leftrightarrow \mu_3\}$. Moreover, note that if we trace over two e.m. tensors at the same time, we end up with a vanishing result. Therefore, we don't have to include terms built with more than one $t_{loc}$ in the decomposition of the correlator.
Proceeding in a manner similar to the previous correlators, we can write the following decomposition
\begin{equation}
	\langle T^{\mu_1\nu_1}T^{\mu_2\nu_2}T^{\mu_3\nu_3}\rangle=\langle t^{\mu_1\nu_1}t^{\mu_2\nu_2}t^{\mu_3\nu_3}\rangle+
	\langle t_{l o c}^{\mu_1\nu_1}t^{\mu_2\nu_2}t^{\mu_3\nu_3}\rangle+
	\langle t^{\mu_1\nu_1}t_{l o c}^{\mu_2\nu_2}t^{\mu_3\nu_3}\rangle+
	\langle t^{\mu_1\nu_1}t^{\mu_2\nu_2}t_{l o c}^{\mu_3\nu_3}\rangle
\end{equation}
with
\begin{equation}
	\langle t^{\mu_1\nu_1}t^{\mu_2\nu_2}t_{l o c}^{\mu_3\nu_3}\rangle=-\frac{16  \, f }{3} \pi^{\mu_3\nu_3}(p_3) \Bigg\{ \bigg[\varepsilon^{\nu_1 \nu_2 p_1 p_2}\bigg((p_1 \cdot p_2) \delta^{\mu_1 \mu_2}- p_1^{\mu_2} p_2^{\mu_1}\bigg) +\left( \mu_1 \leftrightarrow \nu_1 \right) \bigg] +\left( \mu_2 \leftrightarrow \nu_2 \right) \Bigg\}
\end{equation} 
Again, one can derive analogous expression for $\langle t_{l o c}^{\mu_1\nu_1}t^{\mu_2\nu_2}t^{\mu_3\nu_3}\rangle$ and $\langle t^{\mu_1\nu_1}t_{l o c}^{\mu_2\nu_2}t^{\mu_3\nu_3}\rangle$ by exchanging the momenta and indices $\{p_{1}\leftrightarrow p_3,\, \mu_{1}\leftrightarrow \mu_3\}$ and $\{p_{2}\leftrightarrow p_3,\, \mu_{2}\leftrightarrow \mu_3\}$ in the last equation.\\
If we then want to fix the transverse-traceless part of the correlator we need to examine the conformal constraints. The situation here is remarkably different from the $\langle JJT\rangle$ case. We know that $R\tilde{R}$ is a topological anomaly so dilatations are not broken. However in this case special conformal transformation are broken and therefore if we assume an anomaly in the $\langle TTT \rangle$ correlator, we expect that there will also be a non-vanishing transverse-traceless part. This point will be addressed in a separate work. \\

\section{Summary of the results}
We can summarize the result of our analysis before coming to the conclusions. 
We have investigated parity-odd terms in the trace anomaly
\begin{equation}
	\mathcal{A}_{odd}=f_1\, \varepsilon^{\mu \nu \rho \sigma} R_{\alpha \beta \mu \nu} R_{\,\,\,\,\, \rho \sigma}^{\alpha \beta}+f_2\,\varepsilon^{\mu\nu\rho\sigma}F_{\mu\nu}F_{\rho\sigma},
\end{equation}
If we consider the abelian gauge contribution to the trace anomaly, i.e$.$ the Chern-Pontryagin density $F\tilde{F}$, the first term of the anomaly \eqref{eq:defanomduff} can be evaluated by computing the $\langle JJT \rangle$ correlator. We have shown that conformal invariance requires the $\langle JJT\rangle$ correlator to be purely longitudinal. It is possible to fix the longitudinal part of the correlator without breaking the CWIs in order to account for the term in the trace anomaly. In momentum space this is summarized by the following expression
\begin{equation}
	\begin{aligned}
		\langle
		J^{\mu_1}\left(p_1\right) J^{\mu_2}&	\left(p_2\right)
		T^{\mu_3\nu_3}\left(p_3\right) 
		\rangle_{odd}=
		\frac{8}{3}\,f\, \pi^{\mu_3\nu_3}(p_3) \varepsilon^{p_1p_2\mu_1\mu_2}
	\end{aligned}
\end{equation}
 On the other hand, in order to analyze the second term of eq. \eqref{eq:defanomduff} in CFTs, we have examined the $\langle JJO\rangle$ correlator. Since we need to consider $O=T^\mu_\mu$, the conformal dimension of the scalar operator is fixed to $\Delta_3=4$. Remarkably, such value of $\Delta_3$ is the only physical case where the correlator can be different from zero. Furthermore, conformal invariance fixes the $\langle JJO\rangle$ to assume the required anomalous form
\begin{equation}
	\begin{aligned}
		&\langle J^{\mu_1}J^{\mu_2}O_{(\Delta_3=4)}\rangle_{odd} = \mathcal{F}\bigg[ \frac{\delta^2 \mathcal{A}_{odd}(x_3)}{\delta A_{\mu_1}(x_1)\,\, \delta A_{\mu_2}(x_2)} \bigg] 
	\end{aligned}
\end{equation}
where we denoted with $\mathcal{F}[\, \cdot\,]$ the Fourier transform with respect to all the coordinates $(x_1,x_2,x_3)$. We are also ignoring the delta $\delta^4\left(\sum p_i\right)$ appearing after Fourier transforming since it is not included in the definition of the correlator.\\
We have also analyzed the gravitational contribution to the odd-parity trace anomaly, i.e. the Pontryagin density $R\tilde R$.
In this case, the first term in eq. \eqref{eq:defanomduff}, can be evaluated by computing the $\langle TTT\rangle$ correlator.
Contrary to popular belief, such correlator doesn't need to vanish if we allow the presence of a trace term. However, although the regularization and prescriptions adopted by Bonora lead to an anomalous $\langle JJT\rangle$, in the case of the $\langle TTT\rangle$, the correlator vanishes. Indeed, according to Bonora, the gravitational contribution to the trace anomaly comes entirely from the second term of eq. \eqref{eq:defanomduff}. 
Therefore, in this case we don't really need to impose the presence of anomalies for the $\langle TTT\rangle$.
\\
In order to analyze the second term in the anomaly \eqref{eq:defanomduff}, we have considered the $\langle TTO \rangle $ correlator.
Conformal invariance requires the correlator to either vanish or not depending on the value of $\Delta_3$. A non-vanishing case occurs when $\Delta_3=4$, which is exactly what we need for $O=T^\mu_\mu$.
We have shown that the conformal structure assumed by such correlator can exactly explain the odd-parity trace anomaly
\begin{equation}
	\begin{aligned}
		\langle T^{\mu_1\nu_1} T^{\mu_2\nu_2}O_{(\Delta_3=4)}\rangle_{odd} = \mathcal{F}\bigg[ \frac{\delta^2\mathcal{A}_{odd}(x_3)}{\delta g_{\mu_1\nu_1}(x_1)\,\, \delta g_{\mu_2\nu_2}(x_2)} \bigg] 
	\end{aligned}
\end{equation}
The implications for the structure of the effective action in the external gauge and gravity fields can be summarized by rather simple nonlocal functionals. 
Our analysis shows that in terms of the parity-odd contribution to the effective action, in the $\langle JJT\rangle$ case, the
correlator is represented by the term 
\beq
\label{pp1}
\mathcal{S}_{JJT}=f_2\int d^4 x' \sqrt{g(x')}\int d^4 x \sqrt{g(x)}R(x)\Box^{-1}_{x,x'} F\tilde F(x')
\eeq
The entire $\langle JJT\rangle_{odd}$ correlator can be obtained from \eqref{pp1}. Other parity-odd terms my be taken into account by inserting additional stress-energy tensors, constrained by external conservation and CWIs.  \\
In the case of the $\langle TTT\rangle $, the Weyl-variant part of the correlator is generated instead by the functional 
\beq
\mathcal{S}_{TTT}=f_1\int d^4 x' \sqrt{g(x')}\int d^4 x \sqrt{g(x)}R(x)\Box^{-1}_{x,x'} R\tilde R(x').
\eeq
It would be interesting to see what the conformal constraints predict in more general cases. 
However, the pattern for 3-point functions is the usual one, as in the case of the parity-even trace anomalies and ordinary chiral anomalies 
\beq
\mathcal{S}_{AVV}=a_1 \int d^4 x' \sqrt{g(x')}\int d^4 x \sqrt{g(x)}\partial\cdot B\,\Box^{-1}_{x,x'} F\tilde F(x'),
\eeq
with $B$ an axial-vector abelian gauge field.
In both cases, the trace and longitudinal sectors are characterised by the presence of the nonlocal $1/\Box$ interaction. The form is that of a bilinear mixing between a spin 1 (for the chiral current, $\partial\cdot B$) or spin 2 external field (R) and an intermediate scalar or  pseudoscalar interpolating state, times the anomaly contributions ($F\tilde F$ or $R\tilde R$).   As stressed in other related works, these interactions play a role in the cosmological context, in the analysis of the conformal backreaction on gravity, and provide a cosistent basis for nonlocal cosmological models for the dark energy (see for instance \cite{Capozziello:2021bki,Alexander:2009tp}).

\section{Conclusions}
We have studied the conformal constraints on the correlators $\langle JJO\rangle$,
$\langle TTO\rangle$, $\langle JJT\rangle$ and $\langle TTT\rangle$ which are all related to possible parity-odd terms in the conformal anomaly. \\
 In the case of the $\langle JJT\rangle$, the structure of the correlator is limited to a trace sector, while  it is more involved in the case of the $\langle TTT\rangle$. We have shown that in both cases the anomaly constraints can be solved by the exchange of an anomaly pole. The structure of the anomaly actions that account for such correlation functions are very similar to what found in the parity-even cases, discussed in previous works.  \\
We have found that the $\langle JJO\rangle_{odd}$ and $\langle TTO\rangle_{odd}$ can be different from zero in a conformal field theory when the conformal dimension of the scalar operator is $\Delta_3=4$, which is exactly the case for $O=\nabla\cdot J_5$ and $O=T^{\mu}_\mu$. Remarkably, the general expression for the conformal $\langle JJO\rangle$ and $\langle TTO\rangle$ with $\Delta_3=4$ corresponds to functional derivatives of the anomaly $\mathcal{A}_{odd}$
\begin{equation}
	\begin{aligned}
		&\langle J^{\mu_1}J^{\mu_2}O_{(\Delta_3=4)}\rangle = \mathcal{F}\bigg[ \frac{\delta^2\mathcal{A}_{odd}(x_3)}{\delta A_{\mu_1}(x_1)\,\, \delta A_{\mu_2}(x_2)} \bigg] \\
		&\langle T^{\mu_1\nu_1} T^{\mu_2\nu_2}O_{(\Delta_3=4)}\rangle = \mathcal{F}\bigg[ \frac{\delta^2\mathcal{A}_{odd}(x_3)}{\delta g_{\mu_1\nu_1}(x_1)\,\, \delta g_{\mu_2\nu_2}(x_2)} \bigg], \\
	\end{aligned}
\end{equation}
where with $\mathcal{F}[\, \cdot\,]$ we have denoted the Fourier transform with respect to all the coordinates $(x_1,x_2,x_3)$. \\
These correlators may play an important role in cosmology since sources of CP violation in the early universe are certainly welcomed, in order to explain the matter-antimatter asymmetry of our universe. If conformal symmetry is bound to play an important role in the very early universe, we are surely entitled to envision alternative scenarios where the sources of CP violation are directly connected with gravity.\\
The application of such correlators are manifolds and cover a variety of contexts, ranging from holography \cite{Nakayama:2012gu,McFadden:2009fg,Bzowski:2011ab} to the investigation of topological  matter \cite{Chernodub:2021nff,Gooth:2017mbd,Coriano:2022zkj}. It is clear that our analysis are just a first step towards the general study of CP-odd anomalies in CFT and their physical implications. 
\vspace{0.5cm}

\centerline{\bf Acknowledgements}
This work is partially supported by INFN within the Iniziativa Specifica QFT-HEP.  
The work of C. C. and S.L. is funded by the European Union, Next Generation EU, PNRR project "National Centre for HPC, Big Data and Quantum Computing", project code CN00000013 and by INFN iniziativa specifica 
QFT-HEP and QG sky. 
M. M. M. is supported by the European Research Council (ERC) under the European Union as Horizon 2020 research and innovation program (grant agreement No818066) and by Deutsche Forschungsgemeinschaft (DFG, German Research Foundation) under Germany's Excellence Strategy EXC-2181/1 - 390900948 (the Heidelberg STRUCTURES Cluster of Excellence). We thank Marco Bochicchio, Loriano Bonora, Paul Frampton, Raffaele Marotta, Mario Cret\`i and Riccardo Tommasi for discussions. 

\section{Note added after publication}
After the paper was published, we identified an error in the primary and secondary conformal equations of $\langle TTO \rangle$ in section \ref{secEqTTO}. This error has been corrected in the current version of the paper. The previous solution we derived for $\Delta_3 = 4$ remains unchanged and the conclusions and interpretations of the paper are unaffected. However, these corrections led us to discover new non-vanishing solutions for $\langle TTO \rangle$ with different values of $\Delta_3$ (see section \ref{secSolTTO}).
\appendix
\section{The parity-odd $\langle AVV\rangle $ and the linking of different sectors} \label{appendix:AVV}
In this appendix  we briefly summarize the approach followed in the case of the $\langle AVV\rangle $, here denoted as $\langle J J J_5\rangle$, involving two conserved and one anomalous currents, in order to highlight the differences with respect to the $\langle JJT\rangle_{odd} $ case.  For more details we refer to 
\cite{Coriano:2023hts}. In this case we require the following Ward identities

\begin{equation}\label{eq:ppp}
	\begin{aligned}
		& p_{i\mu_i}\,\braket{J^{\mu_1}(p_1)J^{\mu_2}(p_2)J_5^{\mu_3}(p_3)}=0,\qquad\qquad\qquad \qquad \{i=1,2\} \\
		& p_{3\mu_3}\braket{J^{\mu_1}(p_1)J^{\mu_2}(p_2)J_5^{\mu_3}(p_3)}=-8 \, a \, i \, \varepsilon^{p_1p_2\mu_1\mu_2}
	\end{aligned}
\end{equation}
We decompose the currents into the longitudinal and transverse components
\begin{equation}
\begin{aligned}
\label{ex}
&J^{\mu}(p)=j^\mu(p)+j_{ loc}^\mu(p),\\
&J_5^{\mu}(p)=j_5^\mu(p)+j_{5 loc}^\mu(p)\\
\end{aligned}
\end{equation}
where 
\begin{equation}
	\begin{aligned}
		&j^{\mu}=\pi^{\mu}_{\alpha}(p)\,J^{\alpha}(p), &&j_{  loc}^{\mu}(p)=\frac{p^\mu}{p^2}\,p\cdot J(p)\\
		&j_5^{\mu}=\pi^{\mu}_{\alpha}(p)\,J_5^{\alpha}(p),\qquad\qquad &&j_{ 5 loc}^{\mu}(p)=\frac{p^\mu}{p^2}\,p\cdot J_5(p)
	\end{aligned}
\end{equation}
Due to the conservation Ward identities \eqref{eq:ppp} we can write the correlator in the following form
\begin{equation}
	\braket{ J^{\mu_1 }(p_1) J^{\mu_2 } (p_2)J_5^{\mu_3}(p_3)}=\braket{ j^{\mu_1 }(p_1) j^{\mu_2 }(p_2) j_5^{\mu_3}(p_3)}+\braket{j^{\mu_1 }(p_1) j^{\mu_2 }(p_2)\, j_{5 \text { loc }}^{\mu_3}(p_3)}\label{decomp}
\end{equation}
where the second term on the right-hand side is fixed by the anomaly
\begin{equation}\label{eq:longtermAVV}
	\left\langle j^{\mu_1}\left(p_1\right) j^{\mu_2}\left(p_2\right) j_{5 \text { loc }}^{\mu_3}\left(p_3\right)\right\rangle=\frac{p_3^{\mu_3}}{p_3^2} p_{3 \alpha_3}\left\langle j^{\mu_1}\left(p_1\right) j^{\mu_2}\left(p_2\right) J_5^{\alpha_3}\left(p_3\right)\right\rangle=-\frac{8 a i}{p_3^2} \varepsilon^{p_1 p_2 \mu_1 \mu_2} p_3^{\mu_3}
\end{equation}
On the other hand, the transverse component can be formally expressed in terms of two form factors $A_1$ and $A_2$
\begin{equation}
	\begin{aligned}
		\braket{j^{\mu_1}(p_1)j^{\mu_2}(p_2) j^{\mu_3}_5 (p_3)} &=\pi^{\mu_1}_{\alpha_1}(p_1)
		\pi^{\mu_2}_{\alpha_2} (p_2) \pi^{\mu_3}_{\alpha_3}
		\left(p_3\right)\Big[ 	A_1(p_1,p_2,p_3)\,\varepsilon^{p_1p_2\alpha_1\alpha_2}p_1^{\alpha_3} \\
		& \qquad+ 
		A_2(p_1,p_2,p_3)\, \varepsilon^{p_1 \alpha_1\alpha_2\alpha_3}  -
		A_2(p_2,p_1,p_3)\, \varepsilon^{p_2\alpha_1\alpha_2\alpha_3}  
		\Big] \label{decompFin}
	\end{aligned}
\end{equation}
where $	A_1(p_1,p_2,p_3)=-A_1(p_2,p_1,p_3)$ due to the Bose simmetry.\\
The two sectors, the longitudinal one, defined by projecting the entire correlator over the external momenta and the complementary transverse one, are linked together by the CWIs. Indeed, projecting on the transverse sector the special CWIs, we can write

\begin{equation}
	\begin{aligned}
		0=&\pi_{\mu_1}^{\lambda_1}(p_1)
		\pi_{\mu_2}^{\lambda_2} (p_2) \pi_{\mu_3}^{\lambda_3}
		(p_3)\,
		\mathcal{K}^\kappa
		 \bigg[\braket{J^{\mu_1}(p_1)J^{\mu_2}(p_2)J_5^{\mu_3}(p_3)}\bigg]=\\ &\pi_{\mu_1}^{\lambda_1}(p_1)
		\pi_{\mu_2}^{\lambda_2} (p_2) \pi_{\mu_3}^{\lambda_3}
		(p_3) \, \mathcal{K}^\kappa  \bigg[
		\braket{ j^{\mu_1 }(p_1) j^{\mu_2 }(p_2) j_5^{\mu_3}(p_3)}+\braket{j^{\mu_1 }(p_1) j^{\mu_2 }(p_2)\, j_{5 \text { loc }}^{\mu_3}(p_3)}
		\bigg]=\\ &\pi_{\mu_1}^{\lambda_1}(p_1)
		\pi_{\mu_2}^{\lambda_2} (p_2) \pi_{\mu_3}^{\lambda_3}
		(p_3)  \bigg[\mathcal{K}^\kappa 
		\braket{ j^{\mu_1 }(p_1) j^{\mu_2 }(p_2) j_5^{\mu_3}(p_3)} 
		-\frac{16\, a \, i \, (\Delta_3-1)}{p_3^2}
		\,\delta^{\mu_3\kappa} \varepsilon^{\mu_1\mu_2p_1p_2}
		\bigg]
	\end{aligned}
\end{equation}
where in the last line we computed the action of the operator $\mathcal{K}^\kappa $ on the longitudinal part of the correlator. Therefore, the special conformal Ward identities connect the transverse and longitudinal sector of the $\langle AVV\rangle$.
We can then solve such equations together with the dilatations Ward identities in order to determine the structure of the form factor $A_1$ and $A_2$ in the transverse component (see \cite{Coriano:2023hts} for the details). In the end, we have
\begin{equation}
	\left\langle j^{\mu_1}\left(p_1\right) j^{\mu_2}\left(p_2\right) j_5^{\mu_3}\left(p_3\right)\right\rangle=8 i a\, \pi_{\alpha_1}^{\mu_1}\left(p_1\right) \pi_{\alpha_2}^{\mu_2}\left(p_2\right) \pi_{\alpha_3}^{\mu_3}\left(p_3\right)\bigg[p_2^2 I_{3\{1,0,1\}} \varepsilon^{p_1 \alpha_1 \alpha_2 \alpha_3}-p_1^2 I_{3\{0,1,1\}} \varepsilon^{p_2 \alpha_1 \alpha_2 \alpha_3}\bigg]
\end{equation}
Note the appearence of the coefficient of the chiral anomaly in the transverse part of the correlator.
On the other hand, as we have already shown in eq. \eqref{eq:longtermAVV}, the longitudinal part of the correlator is described by a single structure, corresponding to the exchange of an anomaly pole 
in the axial-vector line.\\
The case of the $\langle AVV\rangle$ correlator is remarkably different from the $\langle JJT\rangle_{odd}$. Indeed, for this last correlator, after performing the analysis of the conformal constraints in this paper, we were able to show that the transverse part is not affected by the presence of an anomalous trace term.

\section{3K Integrals} \label{appendix:3kint}
The most general solution of the conformal Ward identities for our correlators can be written in terms of integrals involving a product of three Bessel functions, namely 3K integrals. In this appendix, we will illustrate such integrals and their properties. For a detailed review on the the topic, see \cite{Bzowski:2013sza,Bzowski:2015pba,Bzowski:2015yxv}.
\subsection{Definition and properties}
First, we recall the definition of the general 3K integral
\begin{equation}\label{eq:3kintdef}
	I_{\alpha\left\{\beta_1 \beta_2 \beta_3\right\}}\left(p_1, p_2, p_3\right)=\int d x x^\alpha \prod_{j=1}^3 p_j^{\beta_j} K_{\beta_j}\left(p_j x\right)
\end{equation}
where $K_\nu$ is a modified Bessel function of the second kind 
\begin{equation}
	K_\nu(x)=\frac{\pi}{2} \frac{I_{-\nu}(x)-I_\nu(x)}{\sin (\nu \pi)}, \qquad \nu \notin \mathbb{Z} \qquad\qquad I_\nu(x)=\left(\frac{x}{2}\right)^\nu \sum_{k=0}^{\infty} \frac{1}{\Gamma(k+1) \Gamma(\nu+1+k)}\left(\frac{x}{2}\right)^{2 k}
\end{equation}
with the property
\begin{equation}
	K_n(x)=\lim _{\epsilon \rightarrow 0} K_{n+\epsilon}(x), \quad n \in \mathbb{Z}
\end{equation}
The triple-K integral depends on four parameters: the power $\alpha$ of the integration variable $x$, and the three Bessel function indices $\beta_j$ . The arguments of the 3K integral are magnitudes of momenta $p_j$ with $j = 1, 2, 3$. One can notice the integral is invariant under the exchange $(p_j , \beta_j )\leftrightarrow (p_i , \beta_i )$.
We will also use the reduced version of the 3K integral defined as
\begin{equation}
	J_{N\left\{k_j\right\}}=I_{\frac{d}{2}-1+N\left\{\Delta_j-\frac{d}{2}+k_j\right\}}
\end{equation}
where we introduced the condensed notation $\{k_j \} = \{k_1, k_2, k_3 \}$.
The 3K integral satisfies an equation analogous to the dilatation equation with scaling degree
\begin{equation}
	\text{deg}\left(J_{N\left\{k_j\right\}}\right)=\Delta_t+k_t-2 d-N
\end{equation}
where 
\begin{equation}
	k_t=k_1+k_2+k_3,\qquad\qquad \Delta_t=\Delta_1+\Delta_2+\Delta_3
\end{equation}
From this analysis, it is simple to relate the form factors to the 3K integrals. Indeed, the dilatation Ward identities tell us that a form factor needs to be written as a combination of integrals of the following type
\begin{equation}
	J_{N+k_t,\{k_1,k_2,k_3\}}
\end{equation}
where $N$ is the number of momenta that the form factor multiplies in the decomposition.
Let us now list some useful properties of 3K integrals
\begin{equation}\label{eq:3kprop}
	\begin{aligned}
		&\frac{\partial}{\partial p_n} J_{N\left\{k_j\right\}}  =-p_n J_{N+1\left\{k_j-\delta_{j n}\right\}} \\&
		J_{N\left\{k_j+\delta_{j n}\right\}}  =p_n^2 J_{N\left\{k_j-\delta_{j n}\right\}}+2\left(\Delta_n-\frac{d}{2}+k_n\right) J_{N-1\left\{k_j\right\}} \\&
		\frac{\partial^2}{\partial p_n^2} J_{N\left\{k_j\right\}}  =J_{N+2\left\{k_j\right\}}-2\left(\Delta_n-\frac{d}{2}+k_n-\frac{1}{2}\right) J_{N+1\left\{k_j-\delta_{j n}\right\}}, \\&
		K_n J_{N\left\{k_j\right\}}  \equiv\left(\frac{\partial^2}{\partial p_n^2}+\frac{\left(d+1-2 \Delta_n\right)}{p_n} \frac{\partial}{\partial p_n}\right) J_{N\left\{k_j\right\}}=J_{N+2\left\{k_j\right\}}-2 k_n J_{N+1\left\{k_j-\delta_{j n}\right\}},\\&
		K_{n m} J_{N\left\{k_j\right\}}\equiv (K_n-K_m)J_{N\left\{k_j\right\}} =-2 k_n J_{N+1\left\{k_j-\delta_{j n}\right\}}+2 k_m J_{N+1\left\{k_j-\delta_{j m}\right\}}
	\end{aligned}
\end{equation}

\subsection{Zero momentum limit}
When solving the secondary conformal Ward identities, it may be useful to perform a zero momentum limit.
In this subsection, we review the behaviour of the 3K integrals in the limit $p_3\rightarrow0$. In this limit, the momentum conservation gives
\begin{equation}
	p_{1}^{\mu}=-p_{2}^{\mu} \qquad \Longrightarrow \qquad p_1=p_2 \equiv p
\end{equation}
Assuming that $\alpha>\beta_t-1$ and $\beta_3>0$, we can write
\begin{equation}
	\lim _{p_3 \rightarrow 0} I_{\alpha\left\{\beta_j\right\}}\left(p, p, p_3\right)=p^{\beta_t-\alpha-1} \ell_{\alpha\left\{\beta_j\right\}}
\end{equation}
where 
\small
\begin{equation}
	\ell_{\alpha\left\{\beta_j\right\}}=\frac{2^{\alpha-3} \Gamma\left(\beta_3\right)}{\Gamma\left(\alpha-\beta_3+1\right)} \Gamma\left(\frac{\alpha+\beta_t+1}{2}-\beta_3\right) \Gamma\left(\frac{\alpha-\beta_t+1}{2}+\beta_1\right) \Gamma\left(\frac{\alpha-\beta_t+1}{2}+\beta_2\right) \Gamma\left(\frac{\alpha-\beta_t+1}{2}\right)
\end{equation}
\normalsize
We can derive similar formulas for the case $p_1\rightarrow0$ or $p_2\rightarrow0$ by 
considering the fact that 3K integrals are invariant under the exchange $(p_j , \beta_j )\leftrightarrow (p_i , \beta_i )$.

\subsection{Divergences and regularization}
The 3K integral defined in \eqref{eq:3kintdef} converges when 
\begin{equation}
	\alpha>\sum_{i=1}^3\left|\beta_i\right|-1 \quad ; \quad p_1, p_2, p_3>0
\end{equation}
If $\alpha$ does not satisfy this inequality, the integrals must be defined by an analytic continuation. 
The quantity
\begin{equation}
	\delta \equiv \sum_{j=1}^3\left|\beta_j\right|-1-\alpha
\end{equation}
is the expected degree of divergence.
However, when
\begin{equation}
	\alpha+1 \pm \beta_1 \pm \beta_2 \pm \beta_3=-2 k \quad, \quad k=0,1,2, \dots
\end{equation}
for some non-negative integer $k$ and any choice of the $\pm$ sign, the analytic continuation of the 3K integral generally has poles in the regularization parameter. 
Therefore, if the above condition is satisfied, we need to regularize the integrals. This can be done by shifting the parameters of the 3K integrals as
\begin{equation}
	I_{\alpha\left\{\beta_1, \beta_2, \beta_3\right\}} \rightarrow I_{\tilde{\alpha}\left\{\tilde{\beta}_1, \tilde{\beta}_2, \tilde{\beta}_3\right\}} \quad \Longrightarrow \quad J_{N\left\{k_1, k_2, k_3\right\}} \rightarrow J_{N+u \epsilon\left\{k_1+v_1 \epsilon, k_2+v_2 \epsilon, k_3+v_3 \epsilon\right\}}
\end{equation}
where
\begin{equation}
	\tilde{\alpha}=\alpha+u \epsilon \quad, \quad \tilde{\beta}_1=\beta_1+v_1 \epsilon \quad, \quad \tilde{\beta}_2=\beta_2+v_2 \epsilon \quad, \quad \tilde{\beta}_3=\beta_3+v_3 \epsilon
\end{equation}
or equivalently by considering
\begin{equation}
	d \rightarrow d+2 u \epsilon \quad ; \quad \Delta \rightarrow \Delta_i+\left(u+v_i\right) \epsilon
\end{equation}
In general, the regularisation parameters $u$ and $v_i$ are arbitrary. However, in certain cases, there may be some constraints on them. For simplicity, in this paper we consider the same $v_i=v$ for every $i$.

\subsection{3K integrals and Feynman integrals}
3K integrals are related to Feynman integrals in momentum space. The exact relations were first derived in \cite{Bzowski:2013sza,Bzowski:2015yxv}. Here we briefly show the results. Such expressions have been recently used in order to show the connection between the conformal analysis and the perturbative one for the $\langle AVV\rangle$ correlator \cite{Coriano:2023hts}.\\
Let $K_{d\{\delta_1\delta_2\delta_3\}}$ denote a massless scalar 1-loop 3-point momentum space integral
\begin{equation}
	K_{d\left\{\delta_1 \delta_2 \delta_3\right\}}=\int \frac{\mathrm{d}^d \boldsymbol{k}}{(2 \pi)^d} \frac{1}{k^{2 \delta_3}\left|\boldsymbol{p}_1-\boldsymbol{k}\right|^{2 \delta_2}\left|\boldsymbol{p}_2+\boldsymbol{k}\right|^{2 \delta_1}}
\end{equation}
Any such integral can be expressed in terms of 3K integrals and vice versa. For scalar integrals the relation reads
\begin{equation}
	K_{d\{\delta_1\delta_2\delta_3\}}=\frac{2^{4-\frac{3d}{2}}}{\pi^{\frac{d}{2}}}\times\frac{I_{\frac{d}{2}-1\{\frac{d}{2}+\delta_1-\delta_t,\frac{d}{2}+\delta_2-\delta_t,\frac{d}{2}+\delta_3-\delta_t \}}}{\Gamma(d-\delta_t)\Gamma(\delta_1)\Gamma(\delta_2)\Gamma(\delta_3)}
\end{equation}
where $\delta_t=\delta_1+\delta_2+\delta_3$. Its inverse reads
\begin{equation}
	\begin{aligned}
		& I_{\alpha\left\{\beta_1 \beta_2 \beta_3\right\}}=2^{3 \alpha-1} \pi^{\alpha+1} \Gamma\left(\frac{\alpha+1+\beta_t}{2}\right) \prod_{j=1}^3 \Gamma\left(\frac{\alpha+1+2 \beta_j-\beta_t}{2}\right) \\
		&\hspace{2.5cm} \times K_{2+2 \alpha,\left\{\frac{1}{2}\left(\alpha+1+2 \beta_1-\beta_t\right), \frac{1}{2}\left(\alpha+1+2 \beta_2-\beta_t\right), \frac{1}{2}\left(\alpha+1+2 \beta_3-\beta_t\right)\right\}}
	\end{aligned}
\end{equation}
where $\beta_t=\beta_1+\beta_2+\beta_3$. All tensorial massless 1-loop 3-point momentum-space integrals can also be expressed in terms of a number of 3K integrals when their tensorial structure is resolved by standard methods (for the exact expressions in this case see appendix A.3 of \cite{Bzowski:2013sza}).

\section{Schouten Identities for the $\langle JJT \rangle$} \label{appendix:Schouten}
In this section we will derive the following minimal decomposition used when analyzing the special conformal constraints on the $\langle JJT \rangle$ correlator
\small
\begin{equation} \label{eq:decompJJTspec}
	\begin{aligned}
		0=&\pi_{\mu_1}^{\alpha_1}\left(p_1\right)
		\pi_{\mu_2}^{\alpha_2} \left(p_2\right) 
		\Pi_{\mu_3\nu_3}^{\alpha_3\beta_3} \left(p_3\right) 
		\mathcal{K}^k\left\langle J^{ \mu_1}\left(p_1\right) 
		J^{  \mu_2}\left(p_2\right) T^{\mu_3\nu_3}	\left(p_3\right)\right\rangle =
		\pi_{\mu_1}^{\alpha_1}\left(p_1\right)
		\pi_{\mu_2}^{\alpha_2} \left(p_2\right) 
		\Pi_{\mu_3\nu_3}^{\alpha_3\beta_3} \left(p_3\right) 
		\Biggl[\\ &	
		\left(C_{11}\varepsilon^{p_1\alpha_1\alpha_2\alpha_3}p_1^{\beta_3}
		+C_{12}\varepsilon^{p_2\alpha_1\alpha_2\alpha_3}p_1^{\beta_3}
		+C_{13}\varepsilon^{p_1p_2\alpha_1\alpha_2}p_1^{\alpha_3}p_1^{\beta_3}
		+C_{14}\varepsilon^{p_1p_2\alpha_2\alpha_3}\delta^{\alpha_1\beta_3}\right)p_1^\kappa+\\ &
		\left(C_{21}\varepsilon^{p_1\alpha_1\alpha_2\alpha_3}p_1^{\beta_3}
		+C_{22}\varepsilon^{p_2\alpha_1\alpha_2\alpha_3}p_1^{\beta_3}
		+C_{23}\varepsilon^{p_1p_2\alpha_1\alpha_2}p_1^{\alpha_3}p_1^{\beta_3}
		+C_{24}\varepsilon^{p_1p_2\alpha_2\alpha_3}\delta^{\alpha_1\beta_3}\right)p_2^\kappa+\\&
		C_{31}\varepsilon^{\kappa \mu_1\mu_2\mu_3} p_1^{\nu_3}
		+C_{32}\varepsilon^{p_1\kappa \mu_2\mu_3}\delta^{\mu_1\nu_3}
		+C_{33}\varepsilon^{p_2\kappa \mu_1\mu_3}\delta^{\mu_2\nu_3}+
		C_{34}\varepsilon^{p_1p_2\kappa\mu_3}\delta^{\mu_1\mu_2}p_1^{\nu_3} +\\&
		C_{41}\delta^{\mu_1 \kappa}\varepsilon^{\mu_2\mu_3p_1p_2} p_1^{\nu_3}
		+C_{51}\delta^{\mu_2 \kappa}\varepsilon^{\mu_1\mu_3p_1p_2} p_1^{\nu_3}
		+C_{61}	{\delta^{\mu_3 \kappa}\varepsilon^{p_1\mu_1\mu_2\nu_3}}
		++C_{62}{\delta^{\mu_3 \kappa}\varepsilon^{p_2\mu_1\mu_2\nu_3}}
		\Biggr]
	\end{aligned}
\end{equation}
\normalsize
Such decomposition is obtained first by writing all the possible tensor structures. In particular the tensor related to the primary equations are
\begin{equation}
	\begin{aligned}
		&\varepsilon^{p_1\alpha_1\alpha_2\alpha_3}p_1^{\beta_3}p_1^\kappa,\qquad
		\varepsilon^{p_2\alpha_1\alpha_2\alpha_3}p_1^{\beta_3}p_1^\kappa,\qquad
		\varepsilon^{p_1p_2\alpha_1\alpha_2}p_1^{\alpha_3}p_1^{\beta_3}p_1^\kappa,\qquad
		\varepsilon^{p_1p_2\alpha_2\alpha_3}\delta^{\alpha_1\beta_3}p_1^\kappa\\&
		\cancel{p_2^{\alpha_1}p_1^{\beta_3}\varepsilon^{p_1p_2\alpha_2\alpha_3}p_1^\kappa},\qquad
		\cancel{p_3^{\alpha_2}p_1^{\beta_3}\varepsilon^{p_1p_2\alpha_1\alpha_3}p_1^\kappa},\qquad
		\cancel{\delta^{\beta_3\alpha_2}\varepsilon^{p_1p_2\alpha_1\alpha_3}p_1^\kappa}\\&
		\varepsilon^{p_1\alpha_1\alpha_2\alpha_3}p_1^{\beta_3}p_2^\kappa,\qquad
		\varepsilon^{p_2\alpha_1\alpha_2\alpha_3}p_1^{\beta_3}p_2^\kappa,\qquad
		\varepsilon^{p_1p_2\alpha_1\alpha_2}p_1^{\alpha_3}p_1^{\beta_3}p_2^\kappa,\qquad
		\varepsilon^{p_1p_2\alpha_2\alpha_3}\delta^{\alpha_1\beta_3}p_2^\kappa\\&
		\cancel{p_2^{\alpha_1}p_1^{\beta_3}\varepsilon^{p_1p_2\alpha_2\alpha_3}p_2^\kappa},\qquad
		\cancel{p_3^{\alpha_2}p_1^{\beta_3}\varepsilon^{p_1p_2\alpha_1\alpha_3}p_2^\kappa},\qquad
		\cancel{\delta^{\beta_3\alpha_2}\varepsilon^{p_1p_2\alpha_1\alpha_3}p_2^\kappa}
	\end{aligned}
\end{equation}
while the tensor related to the secondary ones are
\begin{equation}
	\begin{aligned}
		&
		\varepsilon^{\kappa \mu_1\mu_2\mu_3} p_1^{\nu_3},\qquad
		\cancel{\varepsilon^{p_1\kappa \mu_2\mu_3}p_2^{\mu_1} p_1^{\nu_3}},\qquad
		\cancel{\varepsilon^{p_2\kappa \mu_2\mu_3}p_2^{\mu_1} p_1^{\nu_3}},\qquad
		\varepsilon^{p_1\kappa \mu_2\mu_3}\delta^{\mu_1\nu_3},\qquad
		\cancel{\varepsilon^{p_2\kappa \mu_2\mu_3}\delta^{\mu_1\nu_3}}\\
		&
		\cancel{\varepsilon^{p_1\kappa \mu_1\mu_3}p_3^{\mu_2} p_1^{\nu_3}},\qquad
		\cancel{\varepsilon^{p_2\kappa \mu_1\mu_3}p_3^{\mu_2} p_1^{\nu_3}},\qquad
		\cancel{\varepsilon^{p_1\kappa \mu_1\mu_3}\delta^{\mu_2\nu_3}},\qquad
		\varepsilon^{p_2\kappa \mu_1\mu_3}\delta^{\mu_2\nu_3},\qquad
		\cancel{\varepsilon^{p_1\kappa \mu_1\mu_2}p_1^{\mu_3} p_1^{\nu_3}},\\
		&
		\cancel{\varepsilon^{p_2\kappa \mu_1\mu_2}p_1^{\mu_3} p_1^{\nu_3}},\qquad
		\varepsilon^{p_1p_2\kappa\mu_3}\delta^{\mu_1\mu_2}p_1^{\nu_3}  ,\qquad
		\cancel{\varepsilon^{p_1p_2\kappa\mu_3}p_2^{\mu_1}p_3^{\mu_2}p_1^{\nu_3}}  ,\qquad
		\cancel{\varepsilon^{p_1p_2\kappa\mu_3}\delta^{\nu_3\mu_2}p_2^{\mu_1} } ,\qquad
		\cancel{\varepsilon^{p_1p_2\kappa\mu_3}\delta^{\nu_3\mu_1}p_3^{\mu_2}},\\
		&
		\cancel{\varepsilon^{p_1p_2\kappa\mu_2}\delta^{\mu_1\mu_3}p_1^{\nu_3} } ,\qquad
		\cancel{\varepsilon^{p_1p_2\kappa\mu_2}p_2^{\mu_1}p_1^{\mu_3}p_1^{\nu_3}  },\qquad
		\cancel{\varepsilon^{p_1p_2\kappa\mu_1}\delta^{\mu_3\mu_2}p_1^{\nu_3}  },\qquad
		\cancel{\varepsilon^{p_1p_2\kappa\mu_1}p_1^{\nu_3}p_1^{\mu_3}p_3^{\mu_2}},\\
		&\cancel{\delta^{\mu_3 \kappa}\varepsilon^{\mu_1\mu_2p_1p_2} p_1^{\nu_3}},\qquad
		\cancel{\delta^{\mu_3 \kappa}\varepsilon^{\nu_3\mu_1p_1p_2} p_3^{\mu_2}},\qquad
		\cancel{\delta^{\mu_3 \kappa}\varepsilon^{\nu_3\mu_2p_1p_2}p_2^{\mu_1}},\qquad
		{\delta^{\mu_3 \kappa}\varepsilon^{p_1\mu_1\mu_2\nu_3}},\qquad
		{\delta^{\mu_3 \kappa}\varepsilon^{p_2\mu_1\mu_2\nu_3}},\\
		&\delta^{\mu_2 \kappa}\varepsilon^{\mu_1\mu_3p_1p_2} p_1^{\nu_3},\qquad
		\delta^{\mu_1 \kappa}\varepsilon^{\mu_2\mu_3p_1p_2} p_1^{\nu_3}
	\end{aligned}
\end{equation}
Not all of these tensors are independent. Indeed all barred tensors can be rewritten in terms of the non-barred ones thanks to the Schouten identities. Ignoring all the barred tensors, we then end up with the minimal decomposition in eq. \eqref{eq:decompJJTspec}. We will now list all the Schouten identities needed in order to eliminate the barred tensor structures.
The first one we consider is
\begin{equation}
	\varepsilon^{[p_1 p_2 \mu_1\mu_2}\delta^{\mu_3]}_\alpha=0
\end{equation}
which can be contracted with $p_{1\alpha}$, $p_{2\alpha}$, $\delta^\kappa_\alpha$ and $\delta^{\nu_3}_\alpha$ obtaining 
\small
\begin{equation}\label{eq:JJTschoutid1}
	\begin{aligned}
		&\pi_{\mu_1}^{\alpha_1}
		\pi_{\mu_2}^{\alpha_2}
		\Pi_{\mu_3\nu_3}^{\alpha_3\beta_3}
		\biggl(
		\varepsilon^{p_1p_2\mu_1\mu_3}p_3^{\mu_2}
		\biggr)=
		\pi_{\mu_1}^{\alpha_1}
		\pi_{\mu_2}^{\alpha_2} 
		\Pi_{\mu_3\nu_3}^{\alpha_3\beta_3}
		\biggl(-	\frac{p_1^2+p_2^2-p_3^2}{2}\varepsilon^{p_1\mu_1\mu_2\mu_3}-p_1^2\varepsilon^{p_2\mu_1\mu_2\mu_3}-\varepsilon^{p_1p_2\mu_1\mu_2}p_1^{\mu_3}
		\biggr)\\
		&\pi_{\mu_1}^{\alpha_1}
		\pi_{\mu_2}^{\alpha_2}
		\Pi_{\mu_3\nu_3}^{\alpha_3\beta_3}
		\biggl(
		\varepsilon^{p_1p_2\mu_2\mu_3}p_2^{\mu_1}
		\biggr)=
		\pi_{\mu_1}^{\alpha_1}
		\pi_{\mu_2}^{\alpha_2} 
		\Pi_{\mu_3\nu_3}^{\alpha_3\beta_3}
		\biggl(	\frac{p_1^2+p_2^2-p_3^2}{2}\varepsilon^{p_2\mu_1\mu_2\mu_3}
		+p_2^2\varepsilon^{p_1\mu_1\mu_2\mu_3}
		+\varepsilon^{p_1p_2\mu_1\mu_2}p_1^{\mu_3}
		\biggr)\\
		&\pi_{\mu_1}^{\alpha_1}
		\pi_{\mu_2}^{\alpha_2}
		\Pi_{\mu_3\nu_3}^{\alpha_3\beta_3}
		\biggl(
		\varepsilon^{p_1p_2\mu_1\mu_2}\delta^{\kappa\mu_3}
		\biggr)=
		\pi_{\mu_1}^{\alpha_1}
		\pi_{\mu_2}^{\alpha_2} 
		\Pi_{\mu_3\nu_3}^{\alpha_3\beta_3}
		\biggl(	-\varepsilon^{p_2\mu_1\mu_2\mu_3}p_1^\kappa
		+\varepsilon^{p_1\mu_1\mu_2\mu_3}p_2^\kappa
		-\varepsilon^{p_1p_2\mu_2\mu_3}\delta^{\kappa\mu_1}
		+\varepsilon^{p_1p_2\mu_1\mu_3}\delta^{\kappa\mu_2}
		\biggr)\\
		&\pi_{\mu_1}^{\alpha_1}
		\pi_{\mu_2}^{\alpha_2}
		\Pi_{\mu_3\nu_3}^{\alpha_3\beta_3}
		\biggl(
		\varepsilon^{p_1p_2\mu_1\mu_3}\delta^{\mu_2\nu_3}
		\biggr)=
		\pi_{\mu_1}^{\alpha_1}
		\pi_{\mu_2}^{\alpha_2} 
		\Pi_{\mu_3\nu_3}^{\alpha_3\beta_3}
		\biggl(	\varepsilon^{p_1\mu_1\mu_2\mu_3}p_1^{\nu_3}
		+\varepsilon^{p_2\mu_1\mu_2\mu_3}p_1^{\nu_3}
		+\varepsilon^{p_1p_2\mu_2\mu_3}\delta^{\mu_1\nu_3}
		\biggr)
	\end{aligned}
\end{equation}
\normalsize
Then we consider the identities
\begin{equation}
	\begin{aligned}
		&\varepsilon^{[p_1 \mu_1\mu_2\mu_3}\delta^{\kappa]}_\alpha=0,\\
		&\varepsilon^{[p_2 \mu_1\mu_2\mu_3}\delta^{\kappa]}_\alpha=0
	\end{aligned}
\end{equation}
which contracted with $p_{1\alpha}$, $p_{2\alpha}$ and $\delta^{\nu_3}_\alpha$ give
\begin{equation}
	\begin{aligned}
		&\pi_{\mu_1}^{\alpha_1}
		\pi_{\mu_2}^{\alpha_2}
		\Pi_{\mu_3\nu_3}^{\alpha_3\beta_3}
		\biggl(
		\varepsilon^{p_1\kappa\mu_1\mu_3}p_3^{\mu_2}
		\biggr)=
		\pi_{\mu_1}^{\alpha_1}
		\pi_{\mu_2}^{\alpha_2} 
		\Pi_{\mu_3\nu_3}^{\alpha_3\beta_3}
		\biggl(	-p_1^2\varepsilon^{\kappa\mu_1\mu_2\mu_3}+
		\varepsilon^{p_1\mu_1\mu_2\mu_3}p_1^\kappa
		-\varepsilon^{p_1\kappa\mu_1\mu_2}p_1^{\mu_3}
		\biggr)\\
		&\pi_{\mu_1}^{\alpha_1}
		\pi_{\mu_2}^{\alpha_2}
		\Pi_{\mu_3\nu_3}^{\alpha_3\beta_3}
		\biggl(
		\varepsilon^{p_2\kappa\mu_2\mu_3}p_2^{\mu_1}
		\biggr)=
		\pi_{\mu_1}^{\alpha_1}
		\pi_{\mu_2}^{\alpha_2} 
		\Pi_{\mu_3\nu_3}^{\alpha_3\beta_3}
		\biggl(	-p_2^2\varepsilon^{\kappa\mu_1\mu_2\mu_3}
		+\varepsilon^{p_2\kappa\mu_1\mu_2}p_1^{\mu_3}
		+\varepsilon^{p_2\mu_1\mu_2\mu_3}p_2^\kappa
		\biggr)\\
		&\pi_{\mu_1}^{\alpha_1}
		\pi_{\mu_2}^{\alpha_2}
		\Pi_{\mu_3\nu_3}^{\alpha_3\beta_3}
		\biggl(
		\varepsilon^{p_1\kappa\mu_1\mu_2}p_1^{\mu_3}
		\biggr)=
		\pi_{\mu_1}^{\alpha_1}
		\pi_{\mu_2}^{\alpha_2} 
		\Pi_{\mu_3\nu_3}^{\alpha_3\beta_3}
		\biggl(	-\frac{p_1^2+p_2^2-p_3^2}{2}\varepsilon^{\kappa\mu_1\mu_2\mu_3}
		-\varepsilon^{p_1\mu_1\mu_2\mu_3}p_2^\kappa
		+\varepsilon^{p_1\kappa\mu_2\mu_3}p_2^{\mu_1}
		\biggr)\\
		&\pi_{\mu_1}^{\alpha_1}
		\pi_{\mu_2}^{\alpha_2}
		\Pi_{\mu_3\nu_3}^{\alpha_3\beta_3}
		\biggl(
		\varepsilon^{p_2\kappa\mu_1\mu_2}p_1^{\mu_3}
		\biggr)=
		\pi_{\mu_1}^{\alpha_1}
		\pi_{\mu_2}^{\alpha_2} 
		\Pi_{\mu_3\nu_3}^{\alpha_3\beta_3}
		\biggl(	\frac{p_1^2+p_2^2-p_3^2}{2}\varepsilon^{\kappa\mu_1\mu_2\mu_3}
		+\varepsilon^{p_2\mu_1\mu_2\mu_3}p_1^\kappa
		-\varepsilon^{p_2\kappa\mu_1\mu_3}p_3^{\mu_2}
		\biggr)
	\end{aligned}
\end{equation}
\begin{equation}
	\begin{aligned}
		&\pi_{\mu_1}^{\alpha_1}
		\pi_{\mu_2}^{\alpha_2}
		\Pi_{\mu_3\nu_3}^{\alpha_3\beta_3}
		\biggl(
		\varepsilon^{p_1\kappa\mu_1\mu_3}\delta^{\mu_2\nu_3}
		\biggr)=
		\pi_{\mu_1}^{\alpha_1}
		\pi_{\mu_2}^{\alpha_2} 
		\Pi_{\mu_3\nu_3}^{\alpha_3\beta_3}
		\biggl(	\varepsilon^{\kappa\mu_1\mu_2\mu_3}p_1^{\nu_3}
		-\varepsilon^{p_1\mu_1\mu_2\mu_3}\delta^{\kappa\nu_3}
		+\varepsilon^{p_1\kappa\mu_2\mu_3}\delta^{\mu_1\nu_3}
		\biggr)\\
		&\pi_{\mu_1}^{\alpha_1}
		\pi_{\mu_2}^{\alpha_2}
		\Pi_{\mu_3\nu_3}^{\alpha_3\beta_3}
		\biggl(
		\varepsilon^{p_2\kappa\mu_2\mu_3}\delta^{\mu_1\nu_3}
		\biggr)=
		\pi_{\mu_1}^{\alpha_1}
		\pi_{\mu_2}^{\alpha_2} 
		\Pi_{\mu_3\nu_3}^{\alpha_3\beta_3}
		\biggl(	\varepsilon^{\kappa\mu_1\mu_2\mu_3}p_1^{\nu_3}
		+\varepsilon^{p_2\mu_1\mu_2\mu_3}\delta^{\kappa\nu_3}
		+\varepsilon^{p_2\kappa\mu_1\mu_3}\delta^{\mu_2\nu_3}
		\biggr)\\		
	\end{aligned}
\end{equation}
The identity
\begin{equation}
	\begin{aligned} \label{eq:schout072}
		&\varepsilon^{[p_1p_2 \mu_1\mu_2}\delta^{\kappa]}_\alpha=0
	\end{aligned} 
\end{equation}
contracted with $p_{1\alpha}$ and $p_{2\alpha}$ give
\begin{equation}
	\begin{aligned}
		&\pi_{\mu_1}^{\alpha_1}
		\pi_{\mu_2}^{\alpha_2}
		\Pi_{\mu_3\nu_3}^{\alpha_3\beta_3}
		\biggl(
		\varepsilon^{p_1p_2\kappa \mu_1}p_3^{\mu_2}
		\biggr)=
		\pi_{\mu_1}^{\alpha_1}
		\pi_{\mu_2}^{\alpha_2} 
		\Pi_{\mu_3\nu_3}^{\alpha_3\beta_3}
		\biggl(	\frac{p_1^2+p_2^2-p_3^2}{2}\varepsilon^{p_1\kappa \mu_1\mu_2}+p_1^2\varepsilon^{p_2\kappa \mu_1\mu_2}+\varepsilon^{p_1p_2\mu_1\mu_2}p_1^{\kappa}
		\biggr)\\
		&\pi_{\mu_1}^{\alpha_1}
		\pi_{\mu_2}^{\alpha_2}
		\Pi_{\mu_3\nu_3}^{\alpha_3\beta_3}
		\biggl(
		\varepsilon^{p_1p_2\kappa \mu_2}p_2^{\mu_1}
		\biggr)=
		\pi_{\mu_1}^{\alpha_1}
		\pi_{\mu_2}^{\alpha_2} 
		\Pi_{\mu_3\nu_3}^{\alpha_3\beta_3}
		\biggl(	
		-\frac{p_1^2+p_2^2-p_3^2}{2}\varepsilon^{p_2\kappa \mu_1\mu_2}
		-p_2^2\varepsilon^{p_1\kappa\mu_1\mu_2}
		+\varepsilon^{p_1p_2\mu_1\mu_2}p_2^{\kappa}
		\biggr)
	\end{aligned}
\end{equation}
It is worth mentioning that one can contract eq$.$ \eqref{eq:schout072} with $\delta^{\mu_3}_\alpha$ or $\delta^{\nu_3}_\alpha$ obtaining other identities that are not
independent taking in consideration all the identities of this section.
Lastly, we consider the identities
\begin{equation}
	\begin{aligned}
		&\varepsilon^{[p_1p_2 \mu_1 \mu_3}\delta^{\kappa]}_\alpha=0,\\
		&\varepsilon^{[p_1p_2 \mu_2 \mu_3}\delta^{\kappa]}_\alpha=0
	\end{aligned}
\end{equation}
Contracting the first identity with $\delta^{\mu_2}_\alpha$ and the second one with $\delta^{\mu_1}_\alpha$ we obtain 
\begin{equation}
	\begin{aligned}
		&\pi_{\mu_1}^{\alpha_1}
		\pi_{\mu_2}^{\alpha_2}
		\Pi_{\mu_3\nu_3}^{\alpha_3\beta_3}
		\biggl(
		\varepsilon^{p_2\kappa\mu_1\mu_3}p_3^{\mu_2}
		\biggr)=
		\pi_{\mu_1}^{\alpha_1}
		\pi_{\mu_2}^{\alpha_2} 
		\Pi_{\mu_3\nu_3}^{\alpha_3\beta_3}
		\biggl(	\varepsilon^{p_1p_2\mu_1\mu_3}\delta^{\kappa\mu_2}
		-\varepsilon^{p_1p_2\kappa\mu_3}\delta^{\mu_1\mu_2}
		+\varepsilon^{p_1p_2\kappa\mu_1}\delta^{\mu_2\mu_3}
		\biggr)\\
		&\pi_{\mu_1}^{\alpha_1}
		\pi_{\mu_2}^{\alpha_2}
		\Pi_{\mu_3\nu_3}^{\alpha_3\beta_3}
		\biggl(
		\varepsilon^{p_1\kappa\mu_2\mu_3}p_2^{\mu_1}
		\biggr)=
		\pi_{\mu_1}^{\alpha_1}
		\pi_{\mu_2}^{\alpha_2} 
		\Pi_{\mu_3\nu_3}^{\alpha_3\beta_3}
		\biggl(	\varepsilon^{p_1p_2\mu_2\mu_3}\delta^{\kappa\mu_1}
		-\varepsilon^{p_1p_2\kappa\mu_3}\delta^{\mu_1\mu_2}
		+\varepsilon^{p_1p_2\kappa\mu_2}\delta^{\mu_1\mu_3}
		\biggr)
	\end{aligned}
\end{equation}
while, if we contract the first identity with $p_{2\alpha}$ and the second one with $p_{1\alpha}$, we obtain
\small
\begin{equation}
	\begin{aligned}
		\pi_{\mu_1}^{\alpha_1}
		\pi_{\mu_2}^{\alpha_2}
		\Pi_{\mu_3\nu_3}^{\alpha_3\beta_3}&
		\biggl(
		\varepsilon^{p_1p_2\kappa\mu_3}p_2^{\mu_1}
		\biggr)p_3^{\mu_2}p_1^{\nu_3}=\\&
		\pi_{\mu_1}^{\alpha_1}
		\pi_{\mu_2}^{\alpha_2} 
		\Pi_{\mu_3\nu_3}^{\alpha_3\beta_3}
		\biggl(	-p_2^2\varepsilon^{p_1\kappa\mu_1\mu_3}
		-\frac{p_1^2+p_2^2-p_3^2}{2}\varepsilon^{p_2\kappa\mu_1\mu_3}
		-\varepsilon^{p_1p_2\kappa\mu_1}p_1^{\mu_3}
		+\varepsilon^{p_1p_2\mu_1\mu_3}p_2^\kappa 
		\biggr)p_3^{\mu_2}p_1^{\nu_3}\\
		\pi_{\mu_1}^{\alpha_1}
		\pi_{\mu_2}^{\alpha_2}
		\Pi_{\mu_3\nu_3}^{\alpha_3\beta_3}&
		\biggl(
		\varepsilon^{p_1p_2\kappa\mu_3}p_2^{\mu_1}
		\biggr)\delta^{\mu_2\nu_3}=\\&
		\pi_{\mu_1}^{\alpha_1}
		\pi_{\mu_2}^{\alpha_2} 
		\Pi_{\mu_3\nu_3}^{\alpha_3\beta_3}
		\biggl(	
		-p_2^2\varepsilon^{p_1\kappa\mu_1\mu_3}
		-\frac{p_1^2+p_2^2-p_3^2}{2}\varepsilon^{p_2\kappa\mu_1\mu_3}
		-\varepsilon^{p_1p_2\kappa\mu_1}p_1^{\mu_3}
		+\varepsilon^{p_1p_2\mu_1\mu_3}p_2^\kappa
		\biggr)\delta^{\mu_2\nu_3}\\
		\pi_{\mu_1}^{\alpha_1}
		\pi_{\mu_2}^{\alpha_2}
		\Pi_{\mu_3\nu_3}^{\alpha_3\beta_3}&
		\biggl(
		\varepsilon^{p_1p_2\kappa\mu_3}p_3^{\mu_2}
		\biggr)\delta^{\mu_1\nu_3}=\\&
		\pi_{\mu_1}^{\alpha_1}
		\pi_{\mu_2}^{\alpha_2} 
		\Pi_{\mu_3\nu_3}^{\alpha_3\beta_3}
		\biggl(	
		-\frac{p_1^2+p_2^2-p_3^2}{2}\varepsilon^{p_1\kappa\mu_2\mu_3}
		-p_1^2\varepsilon^{p_2\kappa\mu_2\mu_3}
		-\varepsilon^{p_1p_2\mu_2\mu_3}p_1^\kappa 
		-\varepsilon^{p_1p_2\kappa\mu_2}p_1^{\mu_3}
		\biggr)\delta^{\mu_1\nu_3}
	\end{aligned}
\end{equation}
\normalsize
and if we contract the first identity with $p_{1\alpha}$ and the second one with $p_{2\alpha}$ we arrive to
\small
\begin{equation}
	\begin{aligned}
		\pi_{\mu_1}^{\alpha_1}
		\pi_{\mu_2}^{\alpha_2}
		\Pi_{\mu_3\nu_3}^{\alpha_3\beta_3}&
		\biggl(
		\varepsilon^{p_1p_2\kappa\mu_1}p_1^{\mu_3}\delta^{\mu_2\nu_3}
		\biggr)=\\&
		\pi_{\mu_1}^{\alpha_1}
		\pi_{\mu_2}^{\alpha_2} 
		\Pi_{\mu_3\nu_3}^{\alpha_3\beta_3}
		\biggl(	-\frac{p_1^2+p_2^2-p_3^2}{2}\varepsilon^{p_1\kappa\mu_1\mu_3}\delta^{\mu_2\nu_3}
		-p_1^2\varepsilon^{p_2\kappa\mu_1\mu_3}\delta^{\mu_2\nu_3}
		-\varepsilon^{p_1p_2\mu_1\mu_3}p_1^\kappa\delta^{\mu_2\nu_3}
		\biggr)\\
		\pi_{\mu_1}^{\alpha_1}
		\pi_{\mu_2}^{\alpha_2}
		\Pi_{\mu_3\nu_3}^{\alpha_3\beta_3}&
		\biggl(
		\varepsilon^{p_1p_2\kappa\mu_2}p_1^{\mu_3}\delta^{\mu_1\nu_3}
		\biggr)=\\&
		\pi_{\mu_1}^{\alpha_1}
		\pi_{\mu_2}^{\alpha_2} 
		\Pi_{\mu_3\nu_3}^{\alpha_3\beta_3}
		\biggl(	-\frac{p_1^2+p_2^2-p_3^2}{2}\varepsilon^{p_2\kappa\mu_2\mu_3}\delta^{\mu_1\nu_3}
		-p_2^2\varepsilon^{p_1\kappa\mu_2\mu_3}\delta^{\mu_1\nu_3}
		+\varepsilon^{p_1p_2\mu_2\mu_3}p_2^\kappa\delta^{\mu_1\nu_3}
		\biggr)
	\end{aligned}
\end{equation}
\normalsize
If instead we contract both the identities with $\delta^{\nu_3}_\alpha$, we don't obtain new independent relations.\\
In addition to all the identites we have written in this section, we also need to consider all the equations obtained from such relations exchanging $\mu_3\leftrightarrow \nu_3$.


\providecommand{\href}[2]{#2}\begingroup\raggedright\endgroup

\end{document}